\documentclass[aps,prl,floatfix,onecolumn,reprint,amsmath,amssymb,superscriptaddress,longbibliography]{revtex4-1}
\usepackage{amsfonts}
\usepackage{mathrsfs}
\usepackage{amsmath}
\usepackage{color}
\usepackage{natbib}
\usepackage{graphicx}
\usepackage{bm}
\usepackage{amssymb}
\usepackage{mathptmx}
\usepackage{stmaryrd}
\usepackage{xspace}
\usepackage{epstopdf}
\usepackage{dcolumn}
\usepackage{longtable}
\usepackage{multirow}
\usepackage{bbding}
\usepackage[colorlinks=true, letterpaper=true, pdfstartview=FitV, linkcolor=blue, citecolor=blue, urlcolor=blue]{hyperref}

\makeatletter

\newcommand{\Rmnum}[1]{\expandafter\@slowromancap\romannumeral #1@}

\makeatother
\begin{document}

\title{Design of Antiferromagnetic Second-order Band Topology with Rotation Topological Invariants \\ in Two Dimensions}

\author{Fangyang Zhan}
\affiliation{Institute for Structure and Function $\&$ Department of Physics $\&$ Chongqing Key Laboratory for Strongly Coupled Physics, Chongqing University, Chongqing 400044, P. R. China}

\author{Zheng Qin}
\affiliation{Institute for Structure and Function $\&$ Department of Physics $\&$ Chongqing Key Laboratory for Strongly Coupled Physics, Chongqing University, Chongqing 400044, P. R. China}

\author{Dong-Hui Xu}
\affiliation{Institute for Structure and Function $\&$ Department of Physics $\&$ Chongqing Key Laboratory for Strongly Coupled Physics, Chongqing University, Chongqing 400044, P. R. China}
\affiliation{Center of Quantum materials and devices, Chongqing University, Chongqing 400044, P. R. China}

\author{Xiaoyuan Zhou}
\affiliation{Center of Quantum materials and devices, Chongqing University, Chongqing 400044, P. R. China}

\author{Da-Shuai Ma}
\email[]{madason.xin@gmail.com}
\affiliation{Institute for Structure and Function $\&$ Department of Physics $\&$ Chongqing Key Laboratory for Strongly Coupled Physics, Chongqing University, Chongqing 400044, P. R. China}
\affiliation{Center of Quantum materials and devices, Chongqing University, Chongqing 400044, P. R. China}

\author{Rui Wang}
\email[]{rcwang@cqu.edu.cn}
\affiliation{Institute for Structure and Function $\&$ Department of Physics $\&$ Chongqing Key Laboratory for Strongly Coupled Physics, Chongqing University, Chongqing 400044, P. R. China}
\affiliation{Center of Quantum materials and devices, Chongqing University, Chongqing 400044, P. R. China}

\begin{abstract}
The existence of fractionally quantized topological corner states serves as a key indicator for two-dimensional second-order topological insulators (SOTIs), yet has not been experimentally observed in realistic materials. Here, based on effective model analysis and symmetry arguments, we propose a strategy for achieving SOTI phases with in-gap corner states in two dimensional systems with antiferromagnetic (AFM) order. We uncover by a minimum lattice model that the band topology originates from the interplay between intrinsic spin-orbital coupling and interlayer AFM exchange interactions. Using first principles calculations, we show that the 2D AFM SOTI phases can be realized in (MnBi$_2$Te$_4$)(Bi$_2$Te$_3$)$_{m}$ films. Moreover, we demonstrate that the nontrivial corner states are linked to rotation topological invariants under three-fold rotation symmetry $C_3$, resulting in $C_3$-symmetric SOTIs with corner charges fractionally quantized to $\frac{n}{3} \lvert e \rvert $ (mod $e$). Due to the great recent achievements in (MnBi$_2$Te$_4$)(Bi$_2$Te$_3$)$_{m}$ systems, our results providing reliable material candidates for experimentally accessible AFM higher-order band topology would draw intense attentions.
\end{abstract}

\pacs{73.20.At, 71.55.Ak, 74.43.-f}

\keywords{ }

\maketitle

\textit{\textcolor{blue}{Introduction. ---}}
In the past decade, the progress of higher-order topological insulators (HOTIs) has significantly enriched the bulk-boundary correspondence in topological phases of matter~\cite{science.aah6442}.
For a $n$th-order topological insulator (TI) in $d$ spatial dimensions, protected in-gap states are featured at its $(d-n)$ dimensional boundary with $1<n\leq d$.
Among various HOTIs, two-dimensional (2D) second-order topological insulators (SOTIs) that possess a gapped bulk and no gapless edge states have attracted enormous research interest~\cite{nature25777,PhysRevLett.120.026801,PhysRevLett.124.036803,PhysRevLett.119.246401,PhysRevLett.123.256402,PhysRevLett.123.216803,acs.nanolett.9b02719,2021Higher-order,ma2022phononic}.
A 2D SOTI manifests topologically protected $0$-dimensional (0D) corner states, which are usually depicted by fractionally quantized corner charge whose quantization is symmetry protected and can change in discrete jumps at topological phase transitions~\cite{science.aba7604,PhysRevLett.119.246402,PhysRevB.99.245151,PhysRevResearch.1.033074,PhysRevB.103.205123,Saha_2023}.
Currently, although the 2D SOTI phase has been proposed for a long time, almost all of the experimental observations have only been verified in artificial crystal systems, such as photonic systems~\cite{2019waveguides,2019Mohammad,PhysRevLett.122.233903,2021Nonlinear,2020Multipolar}, acoustic systems~\cite{2019acoustic,2019Observation,2019sonic}, and electric circuits~\cite{2018Topolectrical-circuit,PhysRevB.98.201402,PhysRevB.99.020304}.
Despite being the most crucial frontier of topological matter, the 2D SOTI phases in electronic materials have not been found yet in experiments, posing a great obstacle for further studies on higher-order topological physics and related application research.

For a system with magnetic order, the interplay between magnetism and electronic structures would enrich the periodic table of band topology~\cite{Xu2020,Elcoro2021,2022Progress}.
For example, breaking the time-reversal ($T$) symmetry of $\mathbb{Z}_{2}$ TIs by the magnetic substrate is regarded as an efficient way to reach 2D SOTI phases~\cite{PhysRevLett.125.056402,PhysRevLett.124.166804,PhysRevB.106.195303}.
In the field of magnetic topological materials, a major breakthrough is the recent discovery of intrinsically magnetic MnBi$_2$Te$_4$~\cite{science.aax8156,PhysRevX.9.041040,sciadv.aaw5685,PhysRevLett.122.206401,PhysRevLett.122.107202,PhysRevX.11.011003,PhysRevLett.123.096401,acs.nanolett.2c03773,acs.chemmater.8b05017}.
The MnBi$_2$Te$_4$ compound exhibits the van der Waals (vdW) layered structure stacked in a unit of septuple layers (SLs), with intralayer ferromagnetic (FM) and interlayer antiferromagnetic (AFM) exchange couplings, resulting in distinct layer-number dependent magnetic topological phases~\cite{PhysRevLett.122.107202,PhysRevX.11.011003,PhysRevLett.123.096401,acs.nanolett.2c03773,PhysRevB.105.035423}.
Since antiferromagnetism acts as the vanishing net moments and $T$-symmetry breaking, band topology coexisting with the AFM order has recently draw attractive attention \cite{PhysRevB.81.245209,2016Dirac,PhysRevLett.118.106402,nanolett.9b00948,PhysRevLett.122.077203,PhysRevLett.124.066401}, which can launch a brand-new field of AFM topological spintronics \cite{2019Topoantifer}.
It is worth noting that MnBi$_2$Te$_4$ has not only established as the first AFM TI but also constituted the progenitor of AFM topological (MnBi$_2$Te$_4$)(Bi$_2$Te$_3$)$_{m}$ superlattices composed of a MnBi$_2$Te$_4$ SL and a Bi$_2$Te$_3$ quintuple layer (QL),
such as MnBi$_4$Te$_7$ ($m=1$) and MnBi$_6$Te$_{10}$ ($m=2$)~\cite{2019Prediction,sciadv.aax9989,PhysRevB.100.155144,PhysRevX.9.041065,vanderWaals13814,npjTunable,PhysRevB.101.161113,PhysRevMaterials.4.054202}.
While recent advancements have been very encouraging, the exploration of AFM topological phases has been mainly limited to the conventional ($1$th-order) band topology, and especially, the 2D AFM SOTIs are largely unexplored.
Up to now, the only predicted 2D AFM SOTI is monolayer FeSe with checkerboard AFM configuration \cite{22NPJmasskink,22Fragile}, and there is no experimental demonstration yet.
Thus, it is highly desirable to identify more experimentally accessible materials with 2D AFM SOTI features.

In this Letter, using the effective model analysis and symmetry arguments, we propose a strategy for realizing 2D SOTI phases in AFM systems.
Starting from a minimum lattice model of a bilayer system with the AFM configuration, we unveil that higher-order topology originates from the interplay between intrinsic spin-orbital coupling (SOC) and interlayer AFM exchange interactions.
We then demonstrate by first-principles calculations that the 2D AFM SOTI phases can be realized in AFM (MnBi$_2$Te$_4$)(Bi$_2$Te$_3$)$_{m}$ films combined of MnBi$_2$Te$_4$ SL and Bi$_2$Te$_3$ QL building blocks.
Moreover, by employing the concept of rotation topological invariants (RTIs), we uncover that corner states of the 2D AFM SOTIs carry the fractionally quantized corner charge, which is protected by three-fold rotation symmetry $C_3$.
Due to the recent growth and diverse build blocks of (MnBi$_2$Te$_4$)(Bi$_2$Te$_3$)$_{m}$ systems, the emergence of higher-order topological phases in their thin films would offer accessible candidates for studying 2D SOTIs experimentally.

\begin{figure*}
    \centering
     \includegraphics[width=0.72\linewidth]{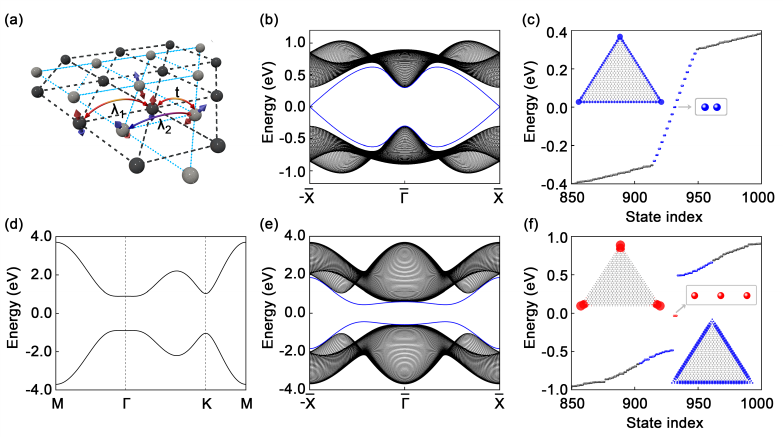}
    \caption{Results of four-band lattice model Eq. (\ref{Eq.1}). (a) Hopping scheme of the symmetry allowed lattice model in MSG 164.88. $t$ describes the spin-conserving hopping for all the nearest neighbors, and $\lambda_{1}$ ($\lambda_{2}$) describes intrinsic spin-orbit coupling that mediates the same (opposite) spins of all the next-nearest neighbors. (b) The edge band structures for a 1D nano-ribbon. (c) The energy spectrum of a triangular nano-disk. In panels (b) and (c), we set the parameters $t=0.3$ eV, $\lambda_{1}=0.2$ eV, and $\lambda_{2}=0$ eV. (d) The bulk band structure for a higher-order topological phase with parameter $\lambda_{2}=0.8$ eV. Panels (e)-(f) same as panels (b)-(c), but including the parameter $\lambda_{2}=0.8$ eV. In panel (f), the insets show three in-gap states (red dots) inside the gaps of both bulk (black dots) and edge (blue dots) states.
     \label{FIG2}}
\end{figure*}

\textit{\textcolor{blue}{Minimum Lattice model of 2D AFM SOTIs. ---}} To elaborate the topological origin of 2D AFM SOTIs, we first construct a minimum lattice model of an AFM bilayer system. Without loss of generality, we consider that this AFM system crystallizes in a trigonal lattice [see Fig.~\ref{FIG2}(a)] and possesses the joint symmetry $PT$, where $P$ denotes the inversion symmetry. We put the single orbital $\vert \varphi \rangle$ at the two $2d$ Wyckoff positions (WPs), \textit{i.e.}, $A:\left(\frac{1}{3},\frac{2}{3},z\right)$ and $B:\left(\frac{2}{3},\frac{1}{3},-z\right)$. In this case, under the spinfull orbital basis $\{ \vert \varphi^A\uparrow\rangle, \vert \varphi^A\downarrow\rangle, \vert \varphi^B\uparrow\rangle, \vert \varphi^B\downarrow\rangle \}$, the symmetry-allowed Hamiltonian can be written as \cite{ZHANG2022108153,PhysRevLett.110.246602,PhysRevB.91.115141}
\begin{equation}\label{Eq.1}
H\left(\boldsymbol{k}\right)  =  \left(\begin{array}{cccc}
f\left(\boldsymbol{k}\right) & g\left(\boldsymbol{k}\right) & h\left(\boldsymbol{k}\right) & 0\\
g\left(\boldsymbol{k}\right)^{*} & -f\left(\boldsymbol{k}\right) & 0 & h\left(\boldsymbol{k}\right)\\
h\left(\boldsymbol{k}\right)^{*} & 0 & -f\left(\boldsymbol{k}\right) & -g\left(\boldsymbol{k}\right)\\
0 & h\left(\boldsymbol{k}\right)^{*} & -g\left(\boldsymbol{k}\right)^{*} & f\left(\boldsymbol{k}\right)
\end{array}\right),
\end{equation}
where $\boldsymbol{k}=(k_1, k_2)$ is the wave vector,
$f\left(\boldsymbol{k}\right)=2 [-\sin(k_1a+k_2a)+\sin k_1a+\sin k_2a] \lambda_{1}$,
$g\left(\boldsymbol{k}\right)= \lambda _{2} [(2+i 2\sqrt{3}/3) \cos (k_1a+k_2a)-  4 i\sqrt{3}/3 \cos k_1a+ (2\sqrt{3}i/3-2) \cos k_2a]$,
and $h\left(\boldsymbol{k}\right)=\left(e^{-i k_1a}+e^{i k_2a}+1\right) t$, with the lattice constance $a$.
As sketched in Fig.~\ref{FIG2}(a), $t$ describes the spin-conserving coupling between the nearest-neighbor sites, while the spin-flip coupling between the nearest-neighbor sites is inhibited completely due to the $PT$ symmetry; $\lambda_{1}$ and $\lambda_{2}$ respectively describe two kinds of intrinsic SOC mediated hoppings~\cite{PhysRevLett.110.246602,PhysRevB.91.115141}.
It is worth noting that $\lambda_{1}$ denotes the spin-conserving next-nearest neighbor (NNN) hopping, which keeps both $P$ symmetry and $T$ symmetry; the NNN spin-flip hopping $\lambda_{2}$ originates from the interlayer coupling due to the interplay of AFM order and intrinsic SOC effect. Thus, a non-zero value of $\lambda_{2}$ breaks the individual $P$ symmetry or $T$ symmetry while maintains the joint symmetry $PT$.

To reveal the key role of couplings originated from the interlayer AFM order, we first ignore the NNN spin-flip hopping $\lambda_{2}$, \textit{i.e.}, we set $t\neq0$, $\lambda_1\neq0$, and $\lambda_2=0$ in the Hamiltonian Eq.~(\ref{Eq.1}). In this case, as shown in Fig.~\ref{FIG2}(b),
there are a pair of helical edge states inside the bulk band gap. These edge bands are gapless, indicating that the bilayer system with $\lambda_{2}=0$ describes a $\mathbb{Z}_{2}$ TI phase. Besides, we also calculate the corresponding energy spectrum of a triangular nano-disk in 0D  geometry with $\lambda_{2}=0$. As shown in Fig.~\ref{FIG2}(c), we can see the pairwise degenerate states (blue dots), which are inside the bulk band gap and explicitly distributed along the edges of triangular nano-disk.

Next, we include the SOC-induced NNN spin-flip hopping (\textit{i.e.}, $\lambda_{2}\neq0$) in the Hamiltonian Eq.~(\ref{Eq.1}). As non-zero value of $\lambda_{2}$ breaks $T$ symmetry, the gapless edge states shown in Fig.~\ref{FIG2}(b) would be gapped (see Fig. S1 in the Supplemental Materials (SM) \cite{SM}). Here, we employ $\lambda_{2}=0.8$ eV, and the corresponding bulk and edge band structures are plotted in Figs.~\ref{FIG2}(d) and \ref{FIG2}(e), respectively. The bands possess a feature of direct band gap at the $\Gamma$ point and gapped edge bands (blue curves) inside the bulk band gap.
We further calculate the energy spectrum of a 0D triangular nano-disk and plot it in Fig.~\ref{FIG2}(f). Remarkably, one can find three in-gap states (red dots) inside the gaps of both bulk (black dots) and edge (blue dots) states. The spatial distribution of these in-gap states are well localized at the three corners [see the inset of Fig.~\ref{FIG2}(f)]. 
Therefore, the minimum lattice model can essentially capture band topology of 2D AFM SOTIs.
From the aspect of lattice model, the interplay between intrinsic SOC and interlayer AFM order leads to non-zero value of $\lambda_2$, which destroys the $T$-symmetry protected Kramers degeneracy of edge states. 
Even so, as the breaking of $T$-symmetry does not affect bulk band inversion, the gapped edge states can give rise to a SOTI phase with 0D corner states~\cite{PhysRevLett.123.256402}. 

\begin{figure}
    \centering
    \includegraphics[width=\linewidth]{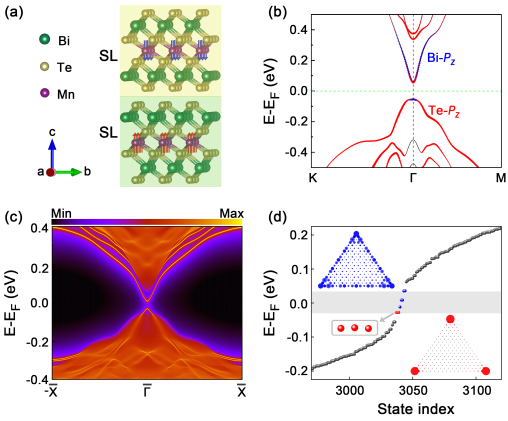}
    \caption{(a) Crystal and magnetic structures of AFM MnBi$_2$Te$_4$ composed of Te-Bi-Te-Mn-Te-Bi-Te SLs. The arrows indicate the magnetic moment directions of the Mn atoms. (b) The orbital-resolved band structures of AFM bilayer MnBi$_2$Te$_4$.  The Te-$p_{z}$ (Bi-$p_{z}$) orbital contribution weight is proportional to the red (blue) curve. (c) The edge states of a semi-finite ribbon of AFM bilayer MnBi$_2$Te$_4$. The gapped edge states can be observed at the $\Gamma$ point. (d) The energy spectrum of $C_3$-symmetric nano-disk of AFM MnBi$_2$Te$_4$  bilayer. The shaded area around the Fermi level represents the bulk band gap. The bulk, edge, and corner states are colored in black, blue, and red, respectively. Corresponding spatial distribution of the edge and corner states are shown in the inserts.
    \label{FIG1}}
\end{figure}

\textit{\textcolor{blue}{SOTI in the AFM MnBi$_2$Te$_4$ bilayer. ---}}
Here, we demonstrate the emergence of SOTI phase established by Eq. (\ref{Eq.1}) can be realized in (MnBi$_2$Te$_4$)(Bi$_2$Te$_3$)$_{m}$ films, in which the presence of AFM order necessitates the existence of an even number of MnBi$_2$Te$_4$ SLs.
Thus, we first focus on electronic properties and band topology of AFM MnBi$_2$Te$_4$ bilayer (\textit{i.e.}, 2 MnBi$_2$Te$_4$ SLs), which can be considered as a prototype of AFM (MnBi$_2$Te$_4$)(Bi$_2$Te$_3$)$_{m}$ films.
As shown in Fig.~\ref{FIG1}(a), the bulk MnBi$_2$Te$_4$ single crystal is composed of Te-Bi-Te-Mn-Te-Bi-Te SLs that are coupled by vdW forces.
Presently, the MnBi$_2$Te$_4$ thin films have been experimental accessible, and the even-number of MnBi$_2$Te$_4$ SLs have compensated AFM magnetic structure with the type-\uppercase\expandafter{\romannumeral3} magnetic space group (MSG) $P3'm'1$ (No. 164.88).
To reveal electronic and topological properties, we carried out first-principles calculations within the framework of density-functional theory~\cite{SMPhysRev.136.B864,SMPhysRev.140.A1133}, and  calculated details were included in the SM~\cite{SM}.
As shown in Fig.~\ref{FIG1}(b), we plot the orbital-resolved electronic band structure of AFM MnBi$_2$Te$_4$ bilayer in the presence of SOC.
The results show that the valence and conduction bands near the Fermi level are dominantly contributed by $p_{z}$ orbitals of Bi and Te atoms.
The calculated band gap at the $\Gamma$ point is about $\sim102$~meV, which is consistent with previously theoretical result of $\sim 107$ meV \cite{PhysRevLett.122.107202}.
More significantly, we can see that a SOC-induced band inversion between Bi-$p_z$ (blue) and Te-$p_z$ (red) orbitals is clearly visible around the $\Gamma$ point, indicating the existence of nontrivial band topology.

To further reveal its nontrivial topological property, we calculate the the local density of states (LDOS) of a semi-finite ribbon of AFM MnBi$_2$Te$_4$ bilayer using the iterative Green's method \cite{SMSancho_1985}. As shown in Fig.~\ref{FIG1}(c), we can observe that there are two edge states with a visible band gap inside the bulk band gap. 
On the other hand, more importantly, gapped edge states coexisted with bulk band inversion imply that it is possible to achieve higher-order band topology~\cite{PhysRevLett.123.256402}.
According to the generalized bulk-boundary correspondence, a 2D HOTI only permits the presence of localized 0D corner states.
Then, we compute the energy spectrum for a triangular nano-disk with 0D geometry, where the $C_3$ symmetry is preserved.
As depicted in Fig.~\ref{FIG1}(d), the calculated energy spectrum 
shows that three in-gap states (colored red) are present.
The spatial distribution of these three in-gap states is well localized at the corners of nano-disk, indicating the emergence of 0D corner states of 2D SOTIs.
Besides, we also display the edge states and their spatial distributions (colored blue) in Fig.~\ref{FIG1}(d). The presence of gapped edge states and in-gap corner states in the AFM MnBi$_2$Te$_4$ bilayer matches with the arguments established by Eq. (\ref{Eq.1}). 

\begin{figure}[b]
    \centering
     \includegraphics[width=\linewidth]{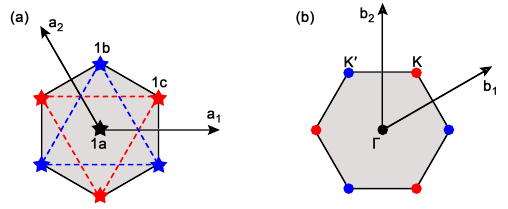}
    \caption{(a) Wyckoff positions  a $C_3$-symmetric system. Here, $1a$, $1b$, and $1c$ are denoted by black, blue, and red stars, respectively. (b) The corresponding Brillouin zones. There are two inequivalent three-fold highly symmetric $\boldsymbol{k}$-points $K$ and $K'$.
    \label{FIG3}}
\end{figure}

\textit{\textcolor{blue}{Rotation topological invariant and corner charge formula. ---}}
Then, we present that the corner states in the AFM MnBi$_2$Te$_4$ bilayer can be further established by the rotation-symmetry protected quantization of fractional corner charge.
The 2D SOTIs protected by $n$-fold rotation-symmetries $C_n$ can be described by nontrivial RTIs and fractionally quantized corner charge \cite{PhysRevB.99.245151,PhysRevResearch.1.033074,PhysRevB.103.205123}.
For a $C_3$ invariant system, there are six RTIs, \textit{i.e.},
$[\Pi_{p}^{\left(3\right)}]=\#\Pi_p^{(3)}-\#\Gamma_p^{(3)}$ ($p=1,2,3$), which represents the difference between the number of states with $C_3$ eigenvalues $e^{\pi (2p-1)i/3}$ at the $\Pi$ and $\Gamma$ points of Brillouin zone (BZ). And, $\Pi$ represents the two unequivalent $C_3$ invariant high-symmetry points $K$ and $K'$.
For an insulator, the number of occupied bands $\nu$ is constant in the whole BZ, \textit{i.e.},
$\sum_{p}\ensuremath{\left[\#\Pi_{p}^{(3)}-\#\Gamma_{p}^{(3)}\right]}\equiv0$.
Thus, to fully describe the rotation-symmetry protected SOTI phases, there are only four linearly independent RTIs,
\begin{equation}\label{Eq.2}
\chi^{(3)}=\{[K_1^{(3)}],[{K'}_1^{(3)}],[K_2^{(3)}],[{K'}_2^{(3)}]\}.
\end{equation}
When the periodic crystal is clipped into a 2D nano-disk with $C_3$ symmetry, the fractionally quantized corner charge is strongly dependent on the WPs occupied by the center of the nano-disk. Specifically, the explicit formulas for the corner
charge in terms of the RTIs reads \cite{PhysRevB.103.205123},
\begin{eqnarray}\label{Eq.3}
Q_{1a}^{(3)} & = & \frac{|e|}{3}\left(n_{1a}^{(\mathrm{ion})}-\nu-[K_{1}^{(3)}]-[K_{2}^{(3)}]-[{K'}_{1}^{(3)}]-[{K'}_{2}^{(3)}]\right)\nonumber \\
 & = & \frac{|e|}{3}\left(n_{1a}^{(\mathrm{ion})}-\nu+[K_{3}^{(3)}]+[{K'}_{3}^{(3)}]\right)\quad(\mathrm{mod}\ e),\nonumber \\
Q_{1b}^{(3)} & = & \frac{|e|}{3}\left(n_{1b}^{(\mathrm{ion})}+[K_{1}^{(3)}]+[{K'}_{2}^{(3)}]\right)\quad(\mathrm{mod}\ e), \\
Q_{1c}^{(3)} & = & \frac{|e|}{3}\left(n_{1c}^{(\mathrm{ion})}+[K_{2}^{(3)}]+[{K'}_{1}^{(3)}]\right)\quad(\mathrm{mod}\ e),\nonumber
\end{eqnarray}
where $n^{(\mathrm{ion})}_{w}$ is ionic charges at WPs $1w$ [see Fig.~\ref{FIG3}(a)].
According to Eq.~(\ref{Eq.3}), three equal fractionally quantized corner charges at $C_3$-related corners may be 0, $\frac{1}{3}\lvert e \rvert$, and $\frac{2}{3}\lvert e \rvert$.

\begin{table}[b]
    \centering
    \renewcommand\arraystretch{1.2}
    \caption{A summary about RTIs and corner charge for the AFM  MnBi$_2$Te$_4$ bilayer and effective lattice model Eq. (\ref{Eq.1}). Here, we take 1$a$ WP as the center of $C_{3}$-symmetric triangle nano-disks.}
    \begin{ruledtabular}
    \begin{tabular}{cccccccc}
       & $n^{(\mathrm{ion})}_{1a}$ & $\nu$ & $[K_{1}^{(3)}]$ & $[{K'}_{1}^{(3)}]$ & $[K_{2}^{(3)}]$ & $[{K'}_{2}^{(3)}]$ & $Q_{1a}^{(3)}$ \\
        \colrule
         AFM-MnBi$_2$Te$_4$   & 22 & 82 & -3 & 6 & -4 & 8 & $\frac{2}{3}\lvert e \rvert$  \\
         Lattice Model   & 0 & 2 & -1 & -1 & 2 & 2 & $\frac{2}{3}\lvert e \rvert$  \\
    \end{tabular}
    \end{ruledtabular}
    \label{Tab. 1}
\end{table}

Now, we apply Eq. (\ref{Eq.3}) to the AFM MnBi$_2$Te$_4$ bilayer and lattice Hamiltonian Eq.~(\ref{Eq.1}), respectively.
The RTIs for the AFM MnBi$_2$Te$_4$  bilayer is $\chi^{(3)}=\{-3,\ 6,\ -4,\ 8\}$, as tabulated in TABLE  \ref{Tab. 1}.
Furthermore, the calculated corner charges of a $C_{3}$-preserved nano-disk that is centered at distinct WPs are $Q_{1a}^{(3)}=Q_{1b}^{(3)}=Q_{1c}^{(3)}=\frac{2}{3}\lvert e \rvert$ (see details in TABLE S\uppercase\expandafter{\romannumeral1} in SM \cite{SM}).
As for the lattice model, we have $\chi^{(3)}_{\mathrm{model}}=\{-1,\ -1,\ 2,\ 2\}$ and fractionally quantized corner charge, \textit{i.e.}, $\frac{2}{3}\lvert e \rvert$.
The result agrees well with that obtained from first-principles calculations. Thus, our minimum lattice model can capture the topological origin of SOTI phase with fractionally quantized corner charge $\frac{2}{3}\lvert e \rvert$ in the AFM MnBi$_2$Te$_4$ bilayer.

\begin{figure}
    \centering
    \includegraphics[width=\linewidth]{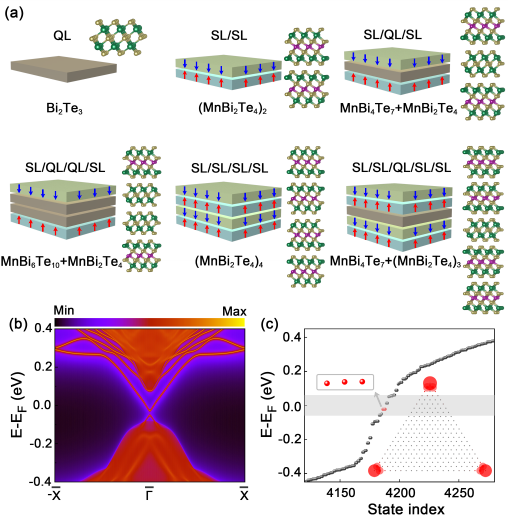}
    \caption{(a) Schematic illustration of crystal structures of several typical AFM (MnBi$_2$Te$_4$)(Bi$_2$Te$_3$)$_{m}$ films, which includes an even number of MnBi$_2$Te$_4$ SLs and an optional number of  Bi$_2$Te$_3$ QLs. (b) The edge states of AFM SL/QL/SL (MnBi$_4$Te$_7$+MnBi$_2$Te$_4$) film. The gapped edge state can also be seen at the $\Gamma$ point. (c) The energy spectra of $C_3$-symmetric triangular nano-disk of AFM SL/QL/SL film. The shaded area represents the bulk band gap. The localized corner states are highlighted in red, and corresponding spatial distributions are shown in the right insert.
    \label{FIG4}}
\end{figure}

\textit{\textcolor{blue}{Design of SOTI phases in AFM (MnBi$_2$Te$_4$)(Bi$_2$Te$_3$)$_{m}$ films. ---}}
Most recently, a great amount of superlatticelike thin films closely related to AFM (MnBi$_2$Te$_4$)(Bi$_2$Te$_3$)$_{m}$ families have been reported to be stable \cite{2019Prediction,sciadv.aax9989,PhysRevB.100.155144,PhysRevX.9.041065,vanderWaals13814,npjTunable,PhysRevB.101.161113,PhysRevMaterials.4.054202}.
In these films, the AFM configuration only requires an even number of MnBi$_2$Te$_4$ SLs, and an optional number of Bi$_2$Te$_3$ QLs can also work as building blocks. Besides, as shown in the SM \cite{SM}, we find that the nonmagnetic Bi$_2$Te$_3$ QL possesses fractionally quantized corner charge (\textit{i.e.}, $\frac{1}{3}\lvert e \rvert$), and its bulk band gap is large enough to $\sim350$~meV.
As a consequence, the diversity of thickness and stacking configurations in these films would provide a reliable avenue to design 2D AFM SOTIs. For instance, the SL/SL configuration represents the MnBi$_2$Te$_4$ bilayer or (MnBi$_2$Te$_4$)$_2$, SL/QL/SL configuration represents MnBi$_2$Te$_4$+MnBi$_4$Te$_7$, SL/QL/QL/SL configuration represents MnBi$_4$Te$_7$+MnBi$_4$Te$_7$ or MnBi$_2$Te$_4$+MnBi$_6$Te$_{10}$, and etc. As shown in Fig.~\ref{FIG4}(a), we illustrate several typical AFM configurations, and the magnetic order of each MnBi$_2$Te$_4$ SL is indicated.
Based on first-principles calculations, we calculate their structure and electronic properties. The calculated results, such as the stacking configurations, bulk band gaps $E_g$, and exchange energies $\Delta E_{\rm ex}=E_{\rm AFM}-E_{\rm FM}$, are concluded in TABLE \ref{Tab. 2}. We can see that all studied films exhibit the AFM ground state with appreciable band gaps.
We further reveal that the corner charges in these AFM films read,
\begin{eqnarray}
Q_{1w}^{(3)} & = & \frac{1}{3}\left(2M+N\right)\lvert e\rvert \quad(\mathrm{mod}\ e),
\end{eqnarray}
where $1w$ represents the WP occupied by the center of the triangular nano-disk, $M$ and $N$ respectively denote the number of 
2 MnBi$_2$Te$_4$ SLs and Bi$_2$Te$_3$ QLs involved in the AFM films. We also list the calculated corner charges in TABLE  \ref{Tab. 2}. It is found that there are always localized corner states with fractionally quantized corner charge, except for the AFM SL/QL/SL (\textit{i.e.}, MnBi$_4$Te$_7$+MnBi$_2$Te$_4$) configuration.
Interestingly, as shown in Figs.~\ref{FIG4}(b) and \ref{FIG4}(c), one can also clearly observe gapped edge states and localized in-gap corner states in the AFM MnBi$_4$Te$_7$+MnBi$_2$Te$_4$.
To understand this, we need turn to its structure, which consists of two building blocks that coupled to each other through vdW interactions, \textit{i.e.}, $M=N=1$.
Thus, the  fractional corner charge in each building block is trivially summarized together, forming one integer corner charge $|e|$.
Notably, although there are in-gap corner states, the absence of non-zero fractional corner charge indicates that filling anomaly in the AFM MnBi$_4$Te$_7$+MnBi$_2$Te$_4$ film is unobservable \cite{PhysRevB.103.205123}.

\begin{table}[t]\footnotesize
    \centering
    \renewcommand\arraystretch{1.2}
    \caption{Properties of several typical AFM (MnBi$_2$Te$_4$)(Bi$_2$Te$_3$)$_{m}$ films, including the stacking configurations, energy gap, exchange energy $\Delta E_{\rm ex}=E_{\rm AFM}-E_{\rm FM}$ between the AFM and FM states, and fractionally quantized corner charges $Q_{1a}^{(3)}$ at $C_3$-related corners.}
    \begin{ruledtabular}
    \begin{tabular}{ccccc}
     Stacking & Compounds & $E_{\rm g}$(meV) & $\Delta E_{\rm ex}$(meV) & $Q_{1a}^{(3)}$ \\
        \colrule
         SL/SL       & (MnBi$_2$Te$_4$)$_2$            & 102 & -0.54 & $\frac{2}{3}\lvert e \rvert$   \\
         SL/QL/SL   & MnBi$_4$Te$_7$+MnBi$_2$Te$_4$         & 123 & -0.15 & 0       \\
         SL/QL/QL/SL   & MnBi$_6$Te$_{10}$+MnBi$_2$Te$_4$         & 54 & -0.01 & $\frac{1}{3}\lvert e \rvert$       \\
         SL/SL/SL/SL     & (MnBi$_2$Te$_4$)$_4$         & 33  & -1.83 & $\frac{1}{3}\lvert e \rvert$  \\
        SL/SL/QL/SL/SL & MnBi$_4$Te$_7$+(MnBi$_2$Te$_4$)$_3$      & 71  & -1.06 & $\frac{2}{3}\lvert e \rvert$   \\
    \end{tabular}
    \end{ruledtabular}
    \label{Tab. 2}
\end{table}

\textit{\textcolor{blue}{Discussion and Summary. ---}}
Summarily, based on the effective lattice model and symmetry arguments, we first prove that the interplay between intrinsic SOC and interlayer AFM exchange interactions can lead to the second-order band topology with in-gap corner modes. We show by first-principles calculations that the emergence of 2D AFM SOTI phases can be realized in (MnBi$_2$Te$_4$)(Bi$_2$Te$_3$)$_{m}$ films with an even number of MnBi$_2$Te$_4$ SLs. In the prototypical AFM MnBi$_2$Te$_4$ bilayer, one can get fractionally quantized corner charge $\frac{2}{3}\lvert e \rvert$ for a $C_3$ symmetry surviving nano-disks that is centered on $1w$ ($w=a,b,c$) WP. Further, we demonstrate the same physics can be generalized to all Mn-Bi-Te superlatticelike films which is in AFM order.
And, the fractionally quantized corner charge of AFM (MnBi$_2$Te$_4$)(Bi$_2$Te$_3$)$_{m}$ films is proved to be closely related to that of the building blocks. 

Since been proposed five years ago, SOTI has been well studied from the aspects of model constructions and  topological classifications.
The realistic material candidates are limited. Comparing with the magnetic substrate induced SOTI \cite{PhysRevLett.125.056402,PhysRevLett.124.166804,PhysRevB.106.195303}, the SOTI phase in the intrinsic AFM (MnBi$_2$Te$_4$)(Bi$_2$Te$_3$)$_{m}$ films would be expected to pose less challenges to experimental realization.
Focusing on experimentally achieved magnetic systems, \textit{i.e.}, AFM (MnBi$_2$Te$_4$)(Bi$_2$Te$_3$)$_{m}$ films, our work
would draw extensive experimental attention.

\textit{\textcolor{blue}{Acknowledgments. ---}}
This work was supported by the National Natural Science Foundation of China (NSFC, Grants No.~12204074, No.~12222402, No.~11974062, and No. 12147102) and the Shenzhen Institute for Quantum Science and Engineering (Grant No. SIQSE202101).


\begin{thebibliography}{72}%
\makeatletter
\providecommand \@ifxundefined [1]{%
 \@ifx{#1\undefined}
}%
\providecommand \@ifnum [1]{%
 \ifnum #1\expandafter \@firstoftwo
 \else \expandafter \@secondoftwo
 \fi
}%
\providecommand \@ifx [1]{%
 \ifx #1\expandafter \@firstoftwo
 \else \expandafter \@secondoftwo
 \fi
}%
\providecommand \natexlab [1]{#1}%
\providecommand \enquote  [1]{``#1''}%
\providecommand \bibnamefont  [1]{#1}%
\providecommand \bibfnamefont [1]{#1}%
\providecommand \citenamefont [1]{#1}%
\providecommand \href@noop [0]{\@secondoftwo}%
\providecommand \href [0]{\begingroup \@sanitize@url \@href}%
\providecommand \@href[1]{\@@startlink{#1}\@@href}%
\providecommand \@@href[1]{\endgroup#1\@@endlink}%
\providecommand \@sanitize@url [0]{\catcode `\\12\catcode `\$12\catcode
  `\&12\catcode `\#12\catcode `\^12\catcode `\_12\catcode `\%12\relax}%
\providecommand \@@startlink[1]{}%
\providecommand \@@endlink[0]{}%
\providecommand \url  [0]{\begingroup\@sanitize@url \@url }%
\providecommand \@url [1]{\endgroup\@href {#1}{\urlprefix }}%
\providecommand \urlprefix  [0]{URL }%
\providecommand \Eprint [0]{\href }%
\providecommand \doibase [0]{https://doi.org/}%
\providecommand \selectlanguage [0]{\@gobble}%
\providecommand \bibinfo  [0]{\@secondoftwo}%
\providecommand \bibfield  [0]{\@secondoftwo}%
\providecommand \translation [1]{[#1]}%
\providecommand \BibitemOpen [0]{}%
\providecommand \bibitemStop [0]{}%
\providecommand \bibitemNoStop [0]{.\EOS\space}%
\providecommand \EOS [0]{\spacefactor3000\relax}%
\providecommand \BibitemShut  [1]{\csname bibitem#1\endcsname}%
\let\auto@bib@innerbib\@empty
\bibitem [{\citenamefont {Benalcazar}\ \emph {et~al.}(2017)\citenamefont
  {Benalcazar}, \citenamefont {Bernevig},\ and\ \citenamefont
  {Hughes}}]{science.aah6442}%
  \BibitemOpen
  \bibfield  {author} {\bibinfo {author} {\bibfnamefont {W.~A.}\ \bibnamefont
  {Benalcazar}}, \bibinfo {author} {\bibfnamefont {B.~A.}\ \bibnamefont
  {Bernevig}},\ and\ \bibinfo {author} {\bibfnamefont {T.~L.}\ \bibnamefont
  {Hughes}},\ }\bibfield  {title} {\bibinfo {title} {Quantized electric
  multipole insulators},\ }\href {https://doi.org/10.1126/science.aah6442}
  {\bibfield  {journal} {\bibinfo  {journal} {Science}\ }\textbf {\bibinfo
  {volume} {357}},\ \bibinfo {pages} {61} (\bibinfo {year} {2017})}\BibitemShut
  {NoStop}%
\bibitem [{\citenamefont {Peterson}\ \emph {et~al.}(2018)\citenamefont
  {Peterson}, \citenamefont {Benalcazar}, \citenamefont {Hughes},\ and\
  \citenamefont {Bahl}}]{nature25777}%
  \BibitemOpen
  \bibfield  {author} {\bibinfo {author} {\bibfnamefont {C.~W.}\ \bibnamefont
  {Peterson}}, \bibinfo {author} {\bibfnamefont {W.~A.}\ \bibnamefont
  {Benalcazar}}, \bibinfo {author} {\bibfnamefont {T.~L.}\ \bibnamefont
  {Hughes}},\ and\ \bibinfo {author} {\bibfnamefont {G.}~\bibnamefont {Bahl}},\
  }\bibfield  {title} {\bibinfo {title} {A quantized microwave quadrupole
  insulator with topologically protected corner states},\ }\href
  {https://doi.org/10.1038/nature25777} {\bibfield  {journal} {\bibinfo
  {journal} {Nature}\ }\textbf {\bibinfo {volume} {555}},\ \bibinfo {pages}
  {346} (\bibinfo {year} {2018})}\BibitemShut {NoStop}%
\bibitem [{\citenamefont {Ezawa}(2018{\natexlab{a}})}]{PhysRevLett.120.026801}%
  \BibitemOpen
  \bibfield  {author} {\bibinfo {author} {\bibfnamefont {M.}~\bibnamefont
  {Ezawa}},\ }\bibfield  {title} {\bibinfo {title} {Higher-order topological
  insulators and semimetals on the breathing kagome and pyrochlore lattices},\
  }\href {https://doi.org/10.1103/PhysRevLett.120.026801} {\bibfield  {journal}
  {\bibinfo  {journal} {Phys. Rev. Lett.}\ }\textbf {\bibinfo {volume} {120}},\
  \bibinfo {pages} {026801} (\bibinfo {year} {2018}{\natexlab{a}})}\BibitemShut
  {NoStop}%
\bibitem [{\citenamefont {Chen}\ \emph
  {et~al.}(2020{\natexlab{a}})\citenamefont {Chen}, \citenamefont {Chen},
  \citenamefont {Gao}, \citenamefont {Zhou},\ and\ \citenamefont
  {Xu}}]{PhysRevLett.124.036803}%
  \BibitemOpen
  \bibfield  {author} {\bibinfo {author} {\bibfnamefont {R.}~\bibnamefont
  {Chen}}, \bibinfo {author} {\bibfnamefont {C.-Z.}\ \bibnamefont {Chen}},
  \bibinfo {author} {\bibfnamefont {J.-H.}\ \bibnamefont {Gao}}, \bibinfo
  {author} {\bibfnamefont {B.}~\bibnamefont {Zhou}},\ and\ \bibinfo {author}
  {\bibfnamefont {D.-H.}\ \bibnamefont {Xu}},\ }\bibfield  {title} {\bibinfo
  {title} {Higher-order topological insulators in quasicrystals},\ }\href
  {https://doi.org/10.1103/PhysRevLett.124.036803} {\bibfield  {journal}
  {\bibinfo  {journal} {Phys. Rev. Lett.}\ }\textbf {\bibinfo {volume} {124}},\
  \bibinfo {pages} {036803} (\bibinfo {year} {2020}{\natexlab{a}})}\BibitemShut
  {NoStop}%
\bibitem [{\citenamefont {Langbehn}\ \emph {et~al.}(2017)\citenamefont
  {Langbehn}, \citenamefont {Peng}, \citenamefont {Trifunovic}, \citenamefont
  {von Oppen},\ and\ \citenamefont {Brouwer}}]{PhysRevLett.119.246401}%
  \BibitemOpen
  \bibfield  {author} {\bibinfo {author} {\bibfnamefont {J.}~\bibnamefont
  {Langbehn}}, \bibinfo {author} {\bibfnamefont {Y.}~\bibnamefont {Peng}},
  \bibinfo {author} {\bibfnamefont {L.}~\bibnamefont {Trifunovic}}, \bibinfo
  {author} {\bibfnamefont {F.}~\bibnamefont {von Oppen}},\ and\ \bibinfo
  {author} {\bibfnamefont {P.~W.}\ \bibnamefont {Brouwer}},\ }\bibfield
  {title} {\bibinfo {title} {Reflection-symmetric second-order topological
  insulators and superconductors},\ }\href
  {https://doi.org/10.1103/PhysRevLett.119.246401} {\bibfield  {journal}
  {\bibinfo  {journal} {Phys. Rev. Lett.}\ }\textbf {\bibinfo {volume} {119}},\
  \bibinfo {pages} {246401} (\bibinfo {year} {2017})}\BibitemShut {NoStop}%
\bibitem [{\citenamefont {Sheng}\ \emph {et~al.}(2019)\citenamefont {Sheng},
  \citenamefont {Chen}, \citenamefont {Liu}, \citenamefont {Chen},
  \citenamefont {Yu}, \citenamefont {Zhao},\ and\ \citenamefont
  {Yang}}]{PhysRevLett.123.256402}%
  \BibitemOpen
  \bibfield  {author} {\bibinfo {author} {\bibfnamefont {X.-L.}\ \bibnamefont
  {Sheng}}, \bibinfo {author} {\bibfnamefont {C.}~\bibnamefont {Chen}},
  \bibinfo {author} {\bibfnamefont {H.}~\bibnamefont {Liu}}, \bibinfo {author}
  {\bibfnamefont {Z.}~\bibnamefont {Chen}}, \bibinfo {author} {\bibfnamefont
  {Z.-M.}\ \bibnamefont {Yu}}, \bibinfo {author} {\bibfnamefont {Y.~X.}\
  \bibnamefont {Zhao}},\ and\ \bibinfo {author} {\bibfnamefont {S.~A.}\
  \bibnamefont {Yang}},\ }\bibfield  {title} {\bibinfo {title} {Two-dimensional
  second-order topological insulator in graphdiyne},\ }\href
  {https://doi.org/10.1103/PhysRevLett.123.256402} {\bibfield  {journal}
  {\bibinfo  {journal} {Phys. Rev. Lett.}\ }\textbf {\bibinfo {volume} {123}},\
  \bibinfo {pages} {256402} (\bibinfo {year} {2019})}\BibitemShut {NoStop}%
\bibitem [{\citenamefont {Park}\ \emph {et~al.}(2019)\citenamefont {Park},
  \citenamefont {Kim}, \citenamefont {Cho},\ and\ \citenamefont
  {Lee}}]{PhysRevLett.123.216803}%
  \BibitemOpen
  \bibfield  {author} {\bibinfo {author} {\bibfnamefont {M.~J.}\ \bibnamefont
  {Park}}, \bibinfo {author} {\bibfnamefont {Y.}~\bibnamefont {Kim}}, \bibinfo
  {author} {\bibfnamefont {G.~Y.}\ \bibnamefont {Cho}},\ and\ \bibinfo {author}
  {\bibfnamefont {S.}~\bibnamefont {Lee}},\ }\bibfield  {title} {\bibinfo
  {title} {Higher-order topological insulator in twisted bilayer graphene},\
  }\href {https://doi.org/10.1103/PhysRevLett.123.216803} {\bibfield  {journal}
  {\bibinfo  {journal} {Phys. Rev. Lett.}\ }\textbf {\bibinfo {volume} {123}},\
  \bibinfo {pages} {216803} (\bibinfo {year} {2019})}\BibitemShut {NoStop}%
\bibitem [{\citenamefont {Liu}\ \emph {et~al.}(2019{\natexlab{a}})\citenamefont
  {Liu}, \citenamefont {Zhao}, \citenamefont {Liu},\ and\ \citenamefont
  {Wang}}]{acs.nanolett.9b02719}%
  \BibitemOpen
  \bibfield  {author} {\bibinfo {author} {\bibfnamefont {B.}~\bibnamefont
  {Liu}}, \bibinfo {author} {\bibfnamefont {G.}~\bibnamefont {Zhao}}, \bibinfo
  {author} {\bibfnamefont {Z.}~\bibnamefont {Liu}},\ and\ \bibinfo {author}
  {\bibfnamefont {Z.~F.}\ \bibnamefont {Wang}},\ }\bibfield  {title} {\bibinfo
  {title} {Two-dimensional quadrupole topological insulator in
  $\gamma$-graphyne},\ }\href {https://doi.org/10.1021/acs.nanolett.9b02719}
  {\bibfield  {journal} {\bibinfo  {journal} {Nano Lett.}\ }\textbf {\bibinfo
  {volume} {19}},\ \bibinfo {pages} {6492} (\bibinfo {year}
  {2019}{\natexlab{a}})}\BibitemShut {NoStop}%
\bibitem [{\citenamefont {Xie}\ \emph {et~al.}(2021)\citenamefont {Xie},
  \citenamefont {Wang}, \citenamefont {Zhang}, \citenamefont {Zhan},
  \citenamefont {Jiang}, \citenamefont {Lu},\ and\ \citenamefont
  {Chen}}]{2021Higher-order}%
  \BibitemOpen
  \bibfield  {author} {\bibinfo {author} {\bibfnamefont {B.}~\bibnamefont
  {Xie}}, \bibinfo {author} {\bibfnamefont {H.-X.}\ \bibnamefont {Wang}},
  \bibinfo {author} {\bibfnamefont {X.}~\bibnamefont {Zhang}}, \bibinfo
  {author} {\bibfnamefont {P.}~\bibnamefont {Zhan}}, \bibinfo {author}
  {\bibfnamefont {J.-H.}\ \bibnamefont {Jiang}}, \bibinfo {author}
  {\bibfnamefont {M.}~\bibnamefont {Lu}},\ and\ \bibinfo {author}
  {\bibfnamefont {Y.}~\bibnamefont {Chen}},\ }\bibfield  {title} {\bibinfo
  {title} {Higher-order band topology},\ }\href
  {https://doi.org/10.1038/s42254-021-00323-4} {\bibfield  {journal} {\bibinfo
  {journal} {Nat. Rev. Phys.}\ }\textbf {\bibinfo {volume} {3}},\ \bibinfo
  {pages} {520} (\bibinfo {year} {2021})}\BibitemShut {NoStop}%
\bibitem [{\citenamefont {Ma}\ \emph {et~al.}()\citenamefont {Ma},
  \citenamefont {Yu}, \citenamefont {Li}, \citenamefont {Zhou},\ and\
  \citenamefont {Wang}}]{ma2022phononic}%
  \BibitemOpen
  \bibfield  {author} {\bibinfo {author} {\bibfnamefont {D.-S.}\ \bibnamefont
  {Ma}}, \bibinfo {author} {\bibfnamefont {K.}~\bibnamefont {Yu}}, \bibinfo
  {author} {\bibfnamefont {X.-P.}\ \bibnamefont {Li}}, \bibinfo {author}
  {\bibfnamefont {X.}~\bibnamefont {Zhou}},\ and\ \bibinfo {author}
  {\bibfnamefont {R.}~\bibnamefont {Wang}},\ }\href@noop {} {\bibinfo {title}
  {Phononic obstructed atomic insulators with robust corner modes}},\ \Eprint
  {https://arxiv.org/abs/2210.14592} {arXiv:2210.14592} \BibitemShut {NoStop}%
\bibitem [{\citenamefont {Peterson}\ \emph {et~al.}(2020)\citenamefont
  {Peterson}, \citenamefont {Li}, \citenamefont {Benalcazar}, \citenamefont
  {Hughes},\ and\ \citenamefont {Bahl}}]{science.aba7604}%
  \BibitemOpen
  \bibfield  {author} {\bibinfo {author} {\bibfnamefont {C.~W.}\ \bibnamefont
  {Peterson}}, \bibinfo {author} {\bibfnamefont {T.}~\bibnamefont {Li}},
  \bibinfo {author} {\bibfnamefont {W.~A.}\ \bibnamefont {Benalcazar}},
  \bibinfo {author} {\bibfnamefont {T.~L.}\ \bibnamefont {Hughes}},\ and\
  \bibinfo {author} {\bibfnamefont {G.}~\bibnamefont {Bahl}},\ }\bibfield
  {title} {\bibinfo {title} {A fractional corner anomaly reveals higher-order
  topology},\ }\href {https://doi.org/10.1126/science.aba7604} {\bibfield
  {journal} {\bibinfo  {journal} {Science}\ }\textbf {\bibinfo {volume}
  {368}},\ \bibinfo {pages} {1114} (\bibinfo {year} {2020})}\BibitemShut
  {NoStop}%
\bibitem [{\citenamefont {Song}\ \emph {et~al.}(2017)\citenamefont {Song},
  \citenamefont {Fang},\ and\ \citenamefont {Fang}}]{PhysRevLett.119.246402}%
  \BibitemOpen
  \bibfield  {author} {\bibinfo {author} {\bibfnamefont {Z.}~\bibnamefont
  {Song}}, \bibinfo {author} {\bibfnamefont {Z.}~\bibnamefont {Fang}},\ and\
  \bibinfo {author} {\bibfnamefont {C.}~\bibnamefont {Fang}},\ }\bibfield
  {title} {\bibinfo {title} {$(d\ensuremath{-}2)$-dimensional edge states of
  rotation symmetry protected topological states},\ }\href
  {https://doi.org/10.1103/PhysRevLett.119.246402} {\bibfield  {journal}
  {\bibinfo  {journal} {Phys. Rev. Lett.}\ }\textbf {\bibinfo {volume} {119}},\
  \bibinfo {pages} {246402} (\bibinfo {year} {2017})}\BibitemShut {NoStop}%
\bibitem [{\citenamefont {Benalcazar}\ \emph {et~al.}(2019)\citenamefont
  {Benalcazar}, \citenamefont {Li},\ and\ \citenamefont
  {Hughes}}]{PhysRevB.99.245151}%
  \BibitemOpen
  \bibfield  {author} {\bibinfo {author} {\bibfnamefont {W.~A.}\ \bibnamefont
  {Benalcazar}}, \bibinfo {author} {\bibfnamefont {T.}~\bibnamefont {Li}},\
  and\ \bibinfo {author} {\bibfnamefont {T.~L.}\ \bibnamefont {Hughes}},\
  }\bibfield  {title} {\bibinfo {title} {Quantization of fractional corner
  charge in ${C}_{n}$-symmetric higher-order topological crystalline
  insulators},\ }\href {https://doi.org/10.1103/PhysRevB.99.245151} {\bibfield
  {journal} {\bibinfo  {journal} {Phys. Rev. B}\ }\textbf {\bibinfo {volume}
  {99}},\ \bibinfo {pages} {245151} (\bibinfo {year} {2019})}\BibitemShut
  {NoStop}%
\bibitem [{\citenamefont {Schindler}\ \emph {et~al.}(2019)\citenamefont
  {Schindler}, \citenamefont {Brzezi\ifmmode~\acute{n}\else \'{n}\fi{}ska},
  \citenamefont {Benalcazar}, \citenamefont {Iraola}, \citenamefont {Bouhon},
  \citenamefont {Tsirkin}, \citenamefont {Vergniory},\ and\ \citenamefont
  {Neupert}}]{PhysRevResearch.1.033074}%
  \BibitemOpen
  \bibfield  {author} {\bibinfo {author} {\bibfnamefont {F.}~\bibnamefont
  {Schindler}}, \bibinfo {author} {\bibfnamefont {M.}~\bibnamefont
  {Brzezi\ifmmode~\acute{n}\else \'{n}\fi{}ska}}, \bibinfo {author}
  {\bibfnamefont {W.~A.}\ \bibnamefont {Benalcazar}}, \bibinfo {author}
  {\bibfnamefont {M.}~\bibnamefont {Iraola}}, \bibinfo {author} {\bibfnamefont
  {A.}~\bibnamefont {Bouhon}}, \bibinfo {author} {\bibfnamefont {S.~S.}\
  \bibnamefont {Tsirkin}}, \bibinfo {author} {\bibfnamefont {M.~G.}\
  \bibnamefont {Vergniory}},\ and\ \bibinfo {author} {\bibfnamefont
  {T.}~\bibnamefont {Neupert}},\ }\bibfield  {title} {\bibinfo {title}
  {Fractional corner charges in spin-orbit coupled crystals},\ }\href
  {https://doi.org/10.1103/PhysRevResearch.1.033074} {\bibfield  {journal}
  {\bibinfo  {journal} {Phys. Rev. Research}\ }\textbf {\bibinfo {volume}
  {1}},\ \bibinfo {pages} {033074} (\bibinfo {year} {2019})}\BibitemShut
  {NoStop}%
\bibitem [{\citenamefont {Takahashi}\ \emph {et~al.}(2021)\citenamefont
  {Takahashi}, \citenamefont {Zhang},\ and\ \citenamefont
  {Murakami}}]{PhysRevB.103.205123}%
  \BibitemOpen
  \bibfield  {author} {\bibinfo {author} {\bibfnamefont {R.}~\bibnamefont
  {Takahashi}}, \bibinfo {author} {\bibfnamefont {T.}~\bibnamefont {Zhang}},\
  and\ \bibinfo {author} {\bibfnamefont {S.}~\bibnamefont {Murakami}},\
  }\bibfield  {title} {\bibinfo {title} {General corner charge formula in
  two-dimensional ${C}_{n}$-symmetric higher-order topological insulators},\
  }\href {https://doi.org/10.1103/PhysRevB.103.205123} {\bibfield  {journal}
  {\bibinfo  {journal} {Phys. Rev. B}\ }\textbf {\bibinfo {volume} {103}},\
  \bibinfo {pages} {205123} (\bibinfo {year} {2021})}\BibitemShut {NoStop}%
\bibitem [{\citenamefont {Saha}\ \emph {et~al.}(2023)\citenamefont {Saha},
  \citenamefont {Nag},\ and\ \citenamefont {Mandal}}]{Saha_2023}%
  \BibitemOpen
  \bibfield  {author} {\bibinfo {author} {\bibfnamefont {S.}~\bibnamefont
  {Saha}}, \bibinfo {author} {\bibfnamefont {T.}~\bibnamefont {Nag}},\ and\
  \bibinfo {author} {\bibfnamefont {S.}~\bibnamefont {Mandal}},\ }\bibfield
  {title} {\bibinfo {title} {Multiple higher-order topological phases with even
  and odd pairs of zero-energy corner modes in a ${C}_{3}$ symmetry broken
  model},\ }\href
  {https://iopscience.iop.org/article/10.1209/0295-5075/acd71a/meta} {\bibfield
   {journal} {\bibinfo  {journal} {EPL-Europhys. Lett.}\ }\textbf {\bibinfo
  {volume} {142}},\ \bibinfo {pages} {56002} (\bibinfo {year}
  {2023})}\BibitemShut {NoStop}%
\bibitem [{\citenamefont {El~Hassan}\ \emph {et~al.}(2019)\citenamefont
  {El~Hassan}, \citenamefont {Kunst}, \citenamefont {Moritz}, \citenamefont
  {Andler}, \citenamefont {Bergholtz},\ and\ \citenamefont
  {Bourennane}}]{2019waveguides}%
  \BibitemOpen
  \bibfield  {author} {\bibinfo {author} {\bibfnamefont {A.}~\bibnamefont
  {El~Hassan}}, \bibinfo {author} {\bibfnamefont {F.~K.}\ \bibnamefont
  {Kunst}}, \bibinfo {author} {\bibfnamefont {A.}~\bibnamefont {Moritz}},
  \bibinfo {author} {\bibfnamefont {G.}~\bibnamefont {Andler}}, \bibinfo
  {author} {\bibfnamefont {E.~J.}\ \bibnamefont {Bergholtz}},\ and\ \bibinfo
  {author} {\bibfnamefont {M.}~\bibnamefont {Bourennane}},\ }\bibfield  {title}
  {\bibinfo {title} {Corner states of light in photonic waveguides},\ }\href
  {https://doi.org/10.1038/s41566-019-0519-y} {\bibfield  {journal} {\bibinfo
  {journal} {Nature Photon.}\ }\textbf {\bibinfo {volume} {13}},\ \bibinfo
  {pages} {697} (\bibinfo {year} {2019})}\BibitemShut {NoStop}%
\bibitem [{\citenamefont {Mittal}\ \emph {et~al.}(2019)\citenamefont {Mittal},
  \citenamefont {Orre}, \citenamefont {Zhu}, \citenamefont {Gorlach},
  \citenamefont {Poddubny},\ and\ \citenamefont {Hafezi}}]{2019Mohammad}%
  \BibitemOpen
  \bibfield  {author} {\bibinfo {author} {\bibfnamefont {S.}~\bibnamefont
  {Mittal}}, \bibinfo {author} {\bibfnamefont {V.~V.}\ \bibnamefont {Orre}},
  \bibinfo {author} {\bibfnamefont {G.}~\bibnamefont {Zhu}}, \bibinfo {author}
  {\bibfnamefont {M.~A.}\ \bibnamefont {Gorlach}}, \bibinfo {author}
  {\bibfnamefont {A.}~\bibnamefont {Poddubny}},\ and\ \bibinfo {author}
  {\bibfnamefont {M.}~\bibnamefont {Hafezi}},\ }\bibfield  {title} {\bibinfo
  {title} {Photonic quadrupole topological phases},\ }\href
  {https://doi.org/10.1038/s41566-019-0452-0} {\bibfield  {journal} {\bibinfo
  {journal} {Nature Photon.}\ }\textbf {\bibinfo {volume} {13}},\ \bibinfo
  {pages} {692} (\bibinfo {year} {2019})}\BibitemShut {NoStop}%
\bibitem [{\citenamefont {Xie}\ \emph {et~al.}(2019)\citenamefont {Xie},
  \citenamefont {Su}, \citenamefont {Wang}, \citenamefont {Su}, \citenamefont
  {Shen}, \citenamefont {Zhan}, \citenamefont {Lu}, \citenamefont {Wang},\ and\
  \citenamefont {Chen}}]{PhysRevLett.122.233903}%
  \BibitemOpen
  \bibfield  {author} {\bibinfo {author} {\bibfnamefont {B.-Y.}\ \bibnamefont
  {Xie}}, \bibinfo {author} {\bibfnamefont {G.-X.}\ \bibnamefont {Su}},
  \bibinfo {author} {\bibfnamefont {H.-F.}\ \bibnamefont {Wang}}, \bibinfo
  {author} {\bibfnamefont {H.}~\bibnamefont {Su}}, \bibinfo {author}
  {\bibfnamefont {X.-P.}\ \bibnamefont {Shen}}, \bibinfo {author}
  {\bibfnamefont {P.}~\bibnamefont {Zhan}}, \bibinfo {author} {\bibfnamefont
  {M.-H.}\ \bibnamefont {Lu}}, \bibinfo {author} {\bibfnamefont {Z.-L.}\
  \bibnamefont {Wang}},\ and\ \bibinfo {author} {\bibfnamefont {Y.-F.}\
  \bibnamefont {Chen}},\ }\bibfield  {title} {\bibinfo {title} {Visualization
  of higher-order topological insulating phases in two-dimensional dielectric
  photonic crystals},\ }\href {https://doi.org/10.1103/PhysRevLett.122.233903}
  {\bibfield  {journal} {\bibinfo  {journal} {Phys. Rev. Lett.}\ }\textbf
  {\bibinfo {volume} {122}},\ \bibinfo {pages} {233903} (\bibinfo {year}
  {2019})}\BibitemShut {NoStop}%
\bibitem [{\citenamefont {Kirsch}\ \emph {et~al.}(2021)\citenamefont {Kirsch},
  \citenamefont {Zhang}, \citenamefont {Kremer}, \citenamefont {Maczewsky},
  \citenamefont {Ivanov}, \citenamefont {Kartashov}, \citenamefont {Torner},
  \citenamefont {Bauer}, \citenamefont {Szameit},\ and\ \citenamefont
  {Heinrich}}]{2021Nonlinear}%
  \BibitemOpen
  \bibfield  {author} {\bibinfo {author} {\bibfnamefont {M.~S.}\ \bibnamefont
  {Kirsch}}, \bibinfo {author} {\bibfnamefont {Y.}~\bibnamefont {Zhang}},
  \bibinfo {author} {\bibfnamefont {M.}~\bibnamefont {Kremer}}, \bibinfo
  {author} {\bibfnamefont {L.~J.}\ \bibnamefont {Maczewsky}}, \bibinfo {author}
  {\bibfnamefont {S.~K.}\ \bibnamefont {Ivanov}}, \bibinfo {author}
  {\bibfnamefont {Y.~V.}\ \bibnamefont {Kartashov}}, \bibinfo {author}
  {\bibfnamefont {L.}~\bibnamefont {Torner}}, \bibinfo {author} {\bibfnamefont
  {D.}~\bibnamefont {Bauer}}, \bibinfo {author} {\bibfnamefont
  {A.}~\bibnamefont {Szameit}},\ and\ \bibinfo {author} {\bibfnamefont
  {M.}~\bibnamefont {Heinrich}},\ }\bibfield  {title} {\bibinfo {title}
  {Nonlinear second-order photonic topological insulators},\ }\href
  {https://doi.org/10.1038/s41567-021-01275-3} {\bibfield  {journal} {\bibinfo
  {journal} {Nat. Phys.}\ }\textbf {\bibinfo {volume} {17}},\ \bibinfo {pages}
  {995} (\bibinfo {year} {2021})}\BibitemShut {NoStop}%
\bibitem [{\citenamefont {Kim}\ \emph {et~al.}(2020)\citenamefont {Kim},
  \citenamefont {Hwang}, \citenamefont {Smirnova}, \citenamefont {Jeong},
  \citenamefont {Kivshar},\ and\ \citenamefont {Park}}]{2020Multipolar}%
  \BibitemOpen
  \bibfield  {author} {\bibinfo {author} {\bibfnamefont {H.-R.}\ \bibnamefont
  {Kim}}, \bibinfo {author} {\bibfnamefont {M.-S.}\ \bibnamefont {Hwang}},
  \bibinfo {author} {\bibfnamefont {D.}~\bibnamefont {Smirnova}}, \bibinfo
  {author} {\bibfnamefont {K.-Y.}\ \bibnamefont {Jeong}}, \bibinfo {author}
  {\bibfnamefont {Y.}~\bibnamefont {Kivshar}},\ and\ \bibinfo {author}
  {\bibfnamefont {H.-G.}\ \bibnamefont {Park}},\ }\bibfield  {title} {\bibinfo
  {title} {Multipolar lasing modes from topological corner states},\ }\href
  {https://doi.org/10.1038/s41467-020-19609-9} {\bibfield  {journal} {\bibinfo
  {journal} {Nat. Commun.}\ }\textbf {\bibinfo {volume} {11}},\ \bibinfo
  {pages} {5758} (\bibinfo {year} {2020})}\BibitemShut {NoStop}%
\bibitem [{\citenamefont {Xue}\ \emph {et~al.}(2019)\citenamefont {Xue},
  \citenamefont {Yang}, \citenamefont {Gao}, \citenamefont {Chong},\ and\
  \citenamefont {Zhang}}]{2019acoustic}%
  \BibitemOpen
  \bibfield  {author} {\bibinfo {author} {\bibfnamefont {H.}~\bibnamefont
  {Xue}}, \bibinfo {author} {\bibfnamefont {Y.}~\bibnamefont {Yang}}, \bibinfo
  {author} {\bibfnamefont {F.}~\bibnamefont {Gao}}, \bibinfo {author}
  {\bibfnamefont {Y.}~\bibnamefont {Chong}},\ and\ \bibinfo {author}
  {\bibfnamefont {B.}~\bibnamefont {Zhang}},\ }\bibfield  {title} {\bibinfo
  {title} {Acoustic higher-order topological insulator on a kagome lattice},\
  }\href {https://doi.org/10.1038/s41563-018-0251-x} {\bibfield  {journal}
  {\bibinfo  {journal} {Nat. Mater.}\ }\textbf {\bibinfo {volume} {18}},\
  \bibinfo {pages} {108} (\bibinfo {year} {2019})}\BibitemShut {NoStop}%
\bibitem [{\citenamefont {Ni}\ \emph {et~al.}(2019)\citenamefont {Ni},
  \citenamefont {Weiner}, \citenamefont {Alu},\ and\ \citenamefont
  {Khanikaev}}]{2019Observation}%
  \BibitemOpen
  \bibfield  {author} {\bibinfo {author} {\bibfnamefont {X.}~\bibnamefont
  {Ni}}, \bibinfo {author} {\bibfnamefont {M.}~\bibnamefont {Weiner}}, \bibinfo
  {author} {\bibfnamefont {A.}~\bibnamefont {Alu}},\ and\ \bibinfo {author}
  {\bibfnamefont {A.~B.}\ \bibnamefont {Khanikaev}},\ }\bibfield  {title}
  {\bibinfo {title} {Observation of higher-order topological acoustic states
  protected by generalized chiral symmetry},\ }\href
  {https://doi.org/10.1038/s41563-018-0252-9} {\bibfield  {journal} {\bibinfo
  {journal} {Nat. Mater.}\ }\textbf {\bibinfo {volume} {18}},\ \bibinfo {pages}
  {113} (\bibinfo {year} {2019})}\BibitemShut {NoStop}%
\bibitem [{\citenamefont {Zhang}\ \emph
  {et~al.}(2019{\natexlab{a}})\citenamefont {Zhang}, \citenamefont {Wang},
  \citenamefont {Lin}, \citenamefont {Tian}, \citenamefont {Xie}, \citenamefont
  {Lu}, \citenamefont {Chen},\ and\ \citenamefont {Jiang}}]{2019sonic}%
  \BibitemOpen
  \bibfield  {author} {\bibinfo {author} {\bibfnamefont {X.}~\bibnamefont
  {Zhang}}, \bibinfo {author} {\bibfnamefont {H.-X.}\ \bibnamefont {Wang}},
  \bibinfo {author} {\bibfnamefont {Z.-K.}\ \bibnamefont {Lin}}, \bibinfo
  {author} {\bibfnamefont {Y.}~\bibnamefont {Tian}}, \bibinfo {author}
  {\bibfnamefont {B.}~\bibnamefont {Xie}}, \bibinfo {author} {\bibfnamefont
  {M.-H.}\ \bibnamefont {Lu}}, \bibinfo {author} {\bibfnamefont {Y.-F.}\
  \bibnamefont {Chen}},\ and\ \bibinfo {author} {\bibfnamefont {J.-H.}\
  \bibnamefont {Jiang}},\ }\bibfield  {title} {\bibinfo {title} {Second-order
  topology and multidimensional topological transitions in sonic crystals},\
  }\href {https://doi.org/10.1038/s41567-019-0472-1} {\bibfield  {journal}
  {\bibinfo  {journal} {Nat. Phys.}\ }\textbf {\bibinfo {volume} {15}},\
  \bibinfo {pages} {582} (\bibinfo {year} {2019}{\natexlab{a}})}\BibitemShut
  {NoStop}%
\bibitem [{\citenamefont {Imhof}\ \emph {et~al.}(2018)\citenamefont {Imhof},
  \citenamefont {Berger}, \citenamefont {Bayer}, \citenamefont {Brehm},
  \citenamefont {Molenkamp}, \citenamefont {Kiessling}, \citenamefont
  {Schindler}, \citenamefont {Lee}, \citenamefont {Greiter}, \citenamefont
  {Neupert},\ and\ \citenamefont {Thomale}}]{2018Topolectrical-circuit}%
  \BibitemOpen
  \bibfield  {author} {\bibinfo {author} {\bibfnamefont {S.}~\bibnamefont
  {Imhof}}, \bibinfo {author} {\bibfnamefont {C.}~\bibnamefont {Berger}},
  \bibinfo {author} {\bibfnamefont {F.}~\bibnamefont {Bayer}}, \bibinfo
  {author} {\bibfnamefont {J.}~\bibnamefont {Brehm}}, \bibinfo {author}
  {\bibfnamefont {L.~W.}\ \bibnamefont {Molenkamp}}, \bibinfo {author}
  {\bibfnamefont {T.}~\bibnamefont {Kiessling}}, \bibinfo {author}
  {\bibfnamefont {F.}~\bibnamefont {Schindler}}, \bibinfo {author}
  {\bibfnamefont {C.~H.}\ \bibnamefont {Lee}}, \bibinfo {author} {\bibfnamefont
  {M.}~\bibnamefont {Greiter}}, \bibinfo {author} {\bibfnamefont
  {T.}~\bibnamefont {Neupert}},\ and\ \bibinfo {author} {\bibfnamefont
  {R.}~\bibnamefont {Thomale}},\ }\bibfield  {title} {\bibinfo {title}
  {Topolectrical-circuit realization of topological corner modes},\ }\href
  {https://doi.org/10.1038/s41567-018-0246-1} {\bibfield  {journal} {\bibinfo
  {journal} {Nat. Phys.}\ }\textbf {\bibinfo {volume} {14}},\ \bibinfo {pages}
  {925} (\bibinfo {year} {2018})}\BibitemShut {NoStop}%
\bibitem [{\citenamefont {Ezawa}(2018{\natexlab{b}})}]{PhysRevB.98.201402}%
  \BibitemOpen
  \bibfield  {author} {\bibinfo {author} {\bibfnamefont {M.}~\bibnamefont
  {Ezawa}},\ }\bibfield  {title} {\bibinfo {title} {Higher-order topological
  electric circuits and topological corner resonance on the breathing kagome
  and pyrochlore lattices},\ }\href
  {https://doi.org/10.1103/PhysRevB.98.201402} {\bibfield  {journal} {\bibinfo
  {journal} {Phys. Rev. B}\ }\textbf {\bibinfo {volume} {98}},\ \bibinfo
  {pages} {201402} (\bibinfo {year} {2018}{\natexlab{b}})}\BibitemShut
  {NoStop}%
\bibitem [{\citenamefont {Serra-Garcia}\ \emph {et~al.}(2019)\citenamefont
  {Serra-Garcia}, \citenamefont {S\"usstrunk},\ and\ \citenamefont
  {Huber}}]{PhysRevB.99.020304}%
  \BibitemOpen
  \bibfield  {author} {\bibinfo {author} {\bibfnamefont {M.}~\bibnamefont
  {Serra-Garcia}}, \bibinfo {author} {\bibfnamefont {R.}~\bibnamefont
  {S\"usstrunk}},\ and\ \bibinfo {author} {\bibfnamefont {S.~D.}\ \bibnamefont
  {Huber}},\ }\bibfield  {title} {\bibinfo {title} {Observation of quadrupole
  transitions and edge mode topology in an {LC} circuit network},\ }\href
  {https://doi.org/10.1103/PhysRevB.99.020304} {\bibfield  {journal} {\bibinfo
  {journal} {Phys. Rev. B}\ }\textbf {\bibinfo {volume} {99}},\ \bibinfo
  {pages} {020304} (\bibinfo {year} {2019})}\BibitemShut {NoStop}%
\bibitem [{\citenamefont {Xu}\ \emph {et~al.}(2020)\citenamefont {Xu},
  \citenamefont {Elcoro}, \citenamefont {Song}, \citenamefont {Wieder},
  \citenamefont {Vergniory}, \citenamefont {Regnault}, \citenamefont {Chen},
  \citenamefont {Felser},\ and\ \citenamefont {Bernevig}}]{Xu2020}%
  \BibitemOpen
  \bibfield  {author} {\bibinfo {author} {\bibfnamefont {Y.}~\bibnamefont
  {Xu}}, \bibinfo {author} {\bibfnamefont {L.}~\bibnamefont {Elcoro}}, \bibinfo
  {author} {\bibfnamefont {Z.-D.}\ \bibnamefont {Song}}, \bibinfo {author}
  {\bibfnamefont {B.~J.}\ \bibnamefont {Wieder}}, \bibinfo {author}
  {\bibfnamefont {M.~G.}\ \bibnamefont {Vergniory}}, \bibinfo {author}
  {\bibfnamefont {N.}~\bibnamefont {Regnault}}, \bibinfo {author}
  {\bibfnamefont {Y.}~\bibnamefont {Chen}}, \bibinfo {author} {\bibfnamefont
  {C.}~\bibnamefont {Felser}},\ and\ \bibinfo {author} {\bibfnamefont {B.~A.}\
  \bibnamefont {Bernevig}},\ }\bibfield  {title} {\bibinfo {title}
  {High-throughput calculations of magnetic topological materials},\ }\href
  {https://doi.org/10.1038/s41586-020-2837-0} {\bibfield  {journal} {\bibinfo
  {journal} {Nature}\ }\textbf {\bibinfo {volume} {586}},\ \bibinfo {pages}
  {702} (\bibinfo {year} {2020})}\BibitemShut {NoStop}%
\bibitem [{\citenamefont {Elcoro}\ \emph {et~al.}(2021)\citenamefont {Elcoro},
  \citenamefont {Wieder}, \citenamefont {Song}, \citenamefont {Xu},
  \citenamefont {Bradlyn},\ and\ \citenamefont {Bernevig}}]{Elcoro2021}%
  \BibitemOpen
  \bibfield  {author} {\bibinfo {author} {\bibfnamefont {L.}~\bibnamefont
  {Elcoro}}, \bibinfo {author} {\bibfnamefont {B.~J.}\ \bibnamefont {Wieder}},
  \bibinfo {author} {\bibfnamefont {Z.}~\bibnamefont {Song}}, \bibinfo {author}
  {\bibfnamefont {Y.}~\bibnamefont {Xu}}, \bibinfo {author} {\bibfnamefont
  {B.}~\bibnamefont {Bradlyn}},\ and\ \bibinfo {author} {\bibfnamefont {B.~A.}\
  \bibnamefont {Bernevig}},\ }\bibfield  {title} {\bibinfo {title} {Magnetic
  topological quantum chemistry},\ }\href
  {https://doi.org/10.1038/s41467-021-26241-8} {\bibfield  {journal} {\bibinfo
  {journal} {Nat. Commun.}\ }\textbf {\bibinfo {volume} {12}},\ \bibinfo
  {pages} {5965} (\bibinfo {year} {2021})}\BibitemShut {NoStop}%
\bibitem [{\citenamefont {Bernevig}\ \emph {et~al.}(2022)\citenamefont
  {Bernevig}, \citenamefont {Felser},\ and\ \citenamefont
  {Beidenkopf}}]{2022Progress}%
  \BibitemOpen
  \bibfield  {author} {\bibinfo {author} {\bibfnamefont {B.~A.}\ \bibnamefont
  {Bernevig}}, \bibinfo {author} {\bibfnamefont {C.}~\bibnamefont {Felser}},\
  and\ \bibinfo {author} {\bibfnamefont {H.}~\bibnamefont {Beidenkopf}},\
  }\bibfield  {title} {\bibinfo {title} {Progress and prospects in magnetic
  topological materials},\ }\href {https://doi.org/10.1038/s41586-021-04105-x}
  {\bibfield  {journal} {\bibinfo  {journal} {Nature}\ }\textbf {\bibinfo
  {volume} {603}},\ \bibinfo {pages} {41} (\bibinfo {year} {2022})}\BibitemShut
  {NoStop}%
\bibitem [{\citenamefont {Chen}\ \emph
  {et~al.}(2020{\natexlab{b}})\citenamefont {Chen}, \citenamefont {Song},
  \citenamefont {Zhao}, \citenamefont {Chen}, \citenamefont {Yu}, \citenamefont
  {Sheng},\ and\ \citenamefont {Yang}}]{PhysRevLett.125.056402}%
  \BibitemOpen
  \bibfield  {author} {\bibinfo {author} {\bibfnamefont {C.}~\bibnamefont
  {Chen}}, \bibinfo {author} {\bibfnamefont {Z.}~\bibnamefont {Song}}, \bibinfo
  {author} {\bibfnamefont {J.-Z.}\ \bibnamefont {Zhao}}, \bibinfo {author}
  {\bibfnamefont {Z.}~\bibnamefont {Chen}}, \bibinfo {author} {\bibfnamefont
  {Z.-M.}\ \bibnamefont {Yu}}, \bibinfo {author} {\bibfnamefont {X.-L.}\
  \bibnamefont {Sheng}},\ and\ \bibinfo {author} {\bibfnamefont {S.~A.}\
  \bibnamefont {Yang}},\ }\bibfield  {title} {\bibinfo {title} {Universal
  approach to magnetic second-order topological insulator},\ }\href
  {https://doi.org/10.1103/PhysRevLett.125.056402} {\bibfield  {journal}
  {\bibinfo  {journal} {Phys. Rev. Lett.}\ }\textbf {\bibinfo {volume} {125}},\
  \bibinfo {pages} {056402} (\bibinfo {year} {2020}{\natexlab{b}})}\BibitemShut
  {NoStop}%
\bibitem [{\citenamefont {Ren}\ \emph {et~al.}(2020)\citenamefont {Ren},
  \citenamefont {Qiao},\ and\ \citenamefont {Niu}}]{PhysRevLett.124.166804}%
  \BibitemOpen
  \bibfield  {author} {\bibinfo {author} {\bibfnamefont {Y.}~\bibnamefont
  {Ren}}, \bibinfo {author} {\bibfnamefont {Z.}~\bibnamefont {Qiao}},\ and\
  \bibinfo {author} {\bibfnamefont {Q.}~\bibnamefont {Niu}},\ }\bibfield
  {title} {\bibinfo {title} {Engineering corner states from two-dimensional
  topological insulators},\ }\href
  {https://doi.org/10.1103/PhysRevLett.124.166804} {\bibfield  {journal}
  {\bibinfo  {journal} {Phys. Rev. Lett.}\ }\textbf {\bibinfo {volume} {124}},\
  \bibinfo {pages} {166804} (\bibinfo {year} {2020})}\BibitemShut {NoStop}%
\bibitem [{\citenamefont {Liu}\ \emph {et~al.}(2022)\citenamefont {Liu},
  \citenamefont {Ren}, \citenamefont {Han}, \citenamefont {Niu},\ and\
  \citenamefont {Qiao}}]{PhysRevB.106.195303}%
  \BibitemOpen
  \bibfield  {author} {\bibinfo {author} {\bibfnamefont {Z.}~\bibnamefont
  {Liu}}, \bibinfo {author} {\bibfnamefont {Y.}~\bibnamefont {Ren}}, \bibinfo
  {author} {\bibfnamefont {Y.}~\bibnamefont {Han}}, \bibinfo {author}
  {\bibfnamefont {Q.}~\bibnamefont {Niu}},\ and\ \bibinfo {author}
  {\bibfnamefont {Z.}~\bibnamefont {Qiao}},\ }\bibfield  {title} {\bibinfo
  {title} {Second-order topological insulator in van der waals heterostructures
  of $\mathrm{CoBr}_{2}/\mathrm{Pt}_{2}\mathrm{HgSe}_{3}/\mathrm{CoBr}_{2}$},\
  }\href {https://doi.org/10.1103/PhysRevB.106.195303} {\bibfield  {journal}
  {\bibinfo  {journal} {Phys. Rev. B}\ }\textbf {\bibinfo {volume} {106}},\
  \bibinfo {pages} {195303} (\bibinfo {year} {2022})}\BibitemShut {NoStop}%
\bibitem [{\citenamefont {Deng}\ \emph {et~al.}(2020)\citenamefont {Deng},
  \citenamefont {Yu}, \citenamefont {Shi}, \citenamefont {Guo}, \citenamefont
  {Xu}, \citenamefont {Wang}, \citenamefont {Chen},\ and\ \citenamefont
  {Zhang}}]{science.aax8156}%
  \BibitemOpen
  \bibfield  {author} {\bibinfo {author} {\bibfnamefont {Y.}~\bibnamefont
  {Deng}}, \bibinfo {author} {\bibfnamefont {Y.}~\bibnamefont {Yu}}, \bibinfo
  {author} {\bibfnamefont {M.~Z.}\ \bibnamefont {Shi}}, \bibinfo {author}
  {\bibfnamefont {Z.}~\bibnamefont {Guo}}, \bibinfo {author} {\bibfnamefont
  {Z.}~\bibnamefont {Xu}}, \bibinfo {author} {\bibfnamefont {J.}~\bibnamefont
  {Wang}}, \bibinfo {author} {\bibfnamefont {X.~H.}\ \bibnamefont {Chen}},\
  and\ \bibinfo {author} {\bibfnamefont {Y.}~\bibnamefont {Zhang}},\ }\bibfield
   {title} {\bibinfo {title} {Quantum anomalous hall effect in intrinsic
  magnetic topological insulator $\mathrm{MnBi}_{2}\mathrm{Te}_{4}$},\ }\href
  {https://doi.org/10.1126/science.aax8156} {\bibfield  {journal} {\bibinfo
  {journal} {Science}\ }\textbf {\bibinfo {volume} {367}},\ \bibinfo {pages}
  {895} (\bibinfo {year} {2020})}\BibitemShut {NoStop}%
\bibitem [{\citenamefont {Chen}\ \emph {et~al.}(2019)\citenamefont {Chen},
  \citenamefont {Xu}, \citenamefont {Li}, \citenamefont {Li}, \citenamefont
  {Wang}, \citenamefont {Zhang}, \citenamefont {Li}, \citenamefont {Wu},
  \citenamefont {Liang}, \citenamefont {Chen}, \citenamefont {Jung},
  \citenamefont {Cacho}, \citenamefont {Mao}, \citenamefont {Liu},
  \citenamefont {Wang}, \citenamefont {Guo}, \citenamefont {Xu}, \citenamefont
  {Liu}, \citenamefont {Yang},\ and\ \citenamefont {Chen}}]{PhysRevX.9.041040}%
  \BibitemOpen
  \bibfield  {author} {\bibinfo {author} {\bibfnamefont {Y.~J.}\ \bibnamefont
  {Chen}}, \bibinfo {author} {\bibfnamefont {L.~X.}\ \bibnamefont {Xu}},
  \bibinfo {author} {\bibfnamefont {J.~H.}\ \bibnamefont {Li}}, \bibinfo
  {author} {\bibfnamefont {Y.~W.}\ \bibnamefont {Li}}, \bibinfo {author}
  {\bibfnamefont {H.~Y.}\ \bibnamefont {Wang}}, \bibinfo {author}
  {\bibfnamefont {C.~F.}\ \bibnamefont {Zhang}}, \bibinfo {author}
  {\bibfnamefont {H.}~\bibnamefont {Li}}, \bibinfo {author} {\bibfnamefont
  {Y.}~\bibnamefont {Wu}}, \bibinfo {author} {\bibfnamefont {A.~J.}\
  \bibnamefont {Liang}}, \bibinfo {author} {\bibfnamefont {C.}~\bibnamefont
  {Chen}}, \bibinfo {author} {\bibfnamefont {S.~W.}\ \bibnamefont {Jung}},
  \bibinfo {author} {\bibfnamefont {C.}~\bibnamefont {Cacho}}, \bibinfo
  {author} {\bibfnamefont {Y.~H.}\ \bibnamefont {Mao}}, \bibinfo {author}
  {\bibfnamefont {S.}~\bibnamefont {Liu}}, \bibinfo {author} {\bibfnamefont
  {M.~X.}\ \bibnamefont {Wang}}, \bibinfo {author} {\bibfnamefont {Y.~F.}\
  \bibnamefont {Guo}}, \bibinfo {author} {\bibfnamefont {Y.}~\bibnamefont
  {Xu}}, \bibinfo {author} {\bibfnamefont {Z.~K.}\ \bibnamefont {Liu}},
  \bibinfo {author} {\bibfnamefont {L.~X.}\ \bibnamefont {Yang}},\ and\
  \bibinfo {author} {\bibfnamefont {Y.~L.}\ \bibnamefont {Chen}},\ }\bibfield
  {title} {\bibinfo {title} {Topological electronic structure and its
  temperature evolution in antiferromagnetic topological insulator
  $\mathrm{MnBi}_{2}\mathrm{Te}_{4}$},\ }\href
  {https://doi.org/10.1103/PhysRevX.9.041040} {\bibfield  {journal} {\bibinfo
  {journal} {Phys. Rev. X}\ }\textbf {\bibinfo {volume} {9}},\ \bibinfo {pages}
  {041040} (\bibinfo {year} {2019})}\BibitemShut {NoStop}%
\bibitem [{\citenamefont {Li}\ \emph {et~al.}(2019)\citenamefont {Li},
  \citenamefont {Li}, \citenamefont {Du}, \citenamefont {Wang}, \citenamefont
  {Gu}, \citenamefont {Zhang}, \citenamefont {He}, \citenamefont {Duan},\ and\
  \citenamefont {Xu}}]{sciadv.aaw5685}%
  \BibitemOpen
  \bibfield  {author} {\bibinfo {author} {\bibfnamefont {J.}~\bibnamefont
  {Li}}, \bibinfo {author} {\bibfnamefont {Y.}~\bibnamefont {Li}}, \bibinfo
  {author} {\bibfnamefont {S.}~\bibnamefont {Du}}, \bibinfo {author}
  {\bibfnamefont {Z.}~\bibnamefont {Wang}}, \bibinfo {author} {\bibfnamefont
  {B.-L.}\ \bibnamefont {Gu}}, \bibinfo {author} {\bibfnamefont {S.-C.}\
  \bibnamefont {Zhang}}, \bibinfo {author} {\bibfnamefont {K.}~\bibnamefont
  {He}}, \bibinfo {author} {\bibfnamefont {W.}~\bibnamefont {Duan}},\ and\
  \bibinfo {author} {\bibfnamefont {Y.}~\bibnamefont {Xu}},\ }\bibfield
  {title} {\bibinfo {title} {Intrinsic magnetic topological insulators in van
  der waals layered $\mathrm{MnBi}_{2}\mathrm{Te}_{4}$-family materials},\
  }\href {https://doi.org/10.1126/sciadv.aaw5685} {\bibfield  {journal}
  {\bibinfo  {journal} {Sci. Adv.}\ }\textbf {\bibinfo {volume} {5}},\ \bibinfo
  {pages} {eaaw5685} (\bibinfo {year} {2019})}\BibitemShut {NoStop}%
\bibitem [{\citenamefont {Zhang}\ \emph
  {et~al.}(2019{\natexlab{b}})\citenamefont {Zhang}, \citenamefont {Shi},
  \citenamefont {Zhu}, \citenamefont {Xing}, \citenamefont {Zhang},\ and\
  \citenamefont {Wang}}]{PhysRevLett.122.206401}%
  \BibitemOpen
  \bibfield  {author} {\bibinfo {author} {\bibfnamefont {D.}~\bibnamefont
  {Zhang}}, \bibinfo {author} {\bibfnamefont {M.}~\bibnamefont {Shi}}, \bibinfo
  {author} {\bibfnamefont {T.}~\bibnamefont {Zhu}}, \bibinfo {author}
  {\bibfnamefont {D.}~\bibnamefont {Xing}}, \bibinfo {author} {\bibfnamefont
  {H.}~\bibnamefont {Zhang}},\ and\ \bibinfo {author} {\bibfnamefont
  {J.}~\bibnamefont {Wang}},\ }\bibfield  {title} {\bibinfo {title}
  {Topological axion states in the magnetic insulator
  $\mathrm{MnBi}_{2}\mathrm{Te}_{4}$ with the quantized magnetoelectric
  effect},\ }\href {https://doi.org/10.1103/PhysRevLett.122.206401} {\bibfield
  {journal} {\bibinfo  {journal} {Phys. Rev. Lett.}\ }\textbf {\bibinfo
  {volume} {122}},\ \bibinfo {pages} {206401} (\bibinfo {year}
  {2019}{\natexlab{b}})}\BibitemShut {NoStop}%
\bibitem [{\citenamefont {Otrokov}\ \emph
  {et~al.}(2019{\natexlab{a}})\citenamefont {Otrokov}, \citenamefont {Rusinov},
  \citenamefont {Blanco-Rey}, \citenamefont {Hoffmann}, \citenamefont
  {Vyazovskaya}, \citenamefont {Eremeev}, \citenamefont {Ernst}, \citenamefont
  {Echenique}, \citenamefont {Arnau},\ and\ \citenamefont
  {Chulkov}}]{PhysRevLett.122.107202}%
  \BibitemOpen
  \bibfield  {author} {\bibinfo {author} {\bibfnamefont {M.~M.}\ \bibnamefont
  {Otrokov}}, \bibinfo {author} {\bibfnamefont {I.~P.}\ \bibnamefont
  {Rusinov}}, \bibinfo {author} {\bibfnamefont {M.}~\bibnamefont {Blanco-Rey}},
  \bibinfo {author} {\bibfnamefont {M.}~\bibnamefont {Hoffmann}}, \bibinfo
  {author} {\bibfnamefont {A.~Y.}\ \bibnamefont {Vyazovskaya}}, \bibinfo
  {author} {\bibfnamefont {S.~V.}\ \bibnamefont {Eremeev}}, \bibinfo {author}
  {\bibfnamefont {A.}~\bibnamefont {Ernst}}, \bibinfo {author} {\bibfnamefont
  {P.~M.}\ \bibnamefont {Echenique}}, \bibinfo {author} {\bibfnamefont
  {A.}~\bibnamefont {Arnau}},\ and\ \bibinfo {author} {\bibfnamefont {E.~V.}\
  \bibnamefont {Chulkov}},\ }\bibfield  {title} {\bibinfo {title} {Unique
  thickness-dependent properties of the van der waals interlayer
  antiferromagnet $\mathrm{MnBi}_{2}\mathrm{Te}_{4}$ films},\ }\href
  {https://doi.org/10.1103/PhysRevLett.122.107202} {\bibfield  {journal}
  {\bibinfo  {journal} {Phys. Rev. Lett.}\ }\textbf {\bibinfo {volume} {122}},\
  \bibinfo {pages} {107202} (\bibinfo {year} {2019}{\natexlab{a}})}\BibitemShut
  {NoStop}%
\bibitem [{\citenamefont {Yang}\ \emph {et~al.}(2021)\citenamefont {Yang},
  \citenamefont {Xu}, \citenamefont {Zhu}, \citenamefont {Niu}, \citenamefont
  {Xu}, \citenamefont {Peng}, \citenamefont {Cheng}, \citenamefont {Jia},
  \citenamefont {Huang}, \citenamefont {Xu}, \citenamefont {Lu},\ and\
  \citenamefont {Ye}}]{PhysRevX.11.011003}%
  \BibitemOpen
  \bibfield  {author} {\bibinfo {author} {\bibfnamefont {S.}~\bibnamefont
  {Yang}}, \bibinfo {author} {\bibfnamefont {X.}~\bibnamefont {Xu}}, \bibinfo
  {author} {\bibfnamefont {Y.}~\bibnamefont {Zhu}}, \bibinfo {author}
  {\bibfnamefont {R.}~\bibnamefont {Niu}}, \bibinfo {author} {\bibfnamefont
  {C.}~\bibnamefont {Xu}}, \bibinfo {author} {\bibfnamefont {Y.}~\bibnamefont
  {Peng}}, \bibinfo {author} {\bibfnamefont {X.}~\bibnamefont {Cheng}},
  \bibinfo {author} {\bibfnamefont {X.}~\bibnamefont {Jia}}, \bibinfo {author}
  {\bibfnamefont {Y.}~\bibnamefont {Huang}}, \bibinfo {author} {\bibfnamefont
  {X.}~\bibnamefont {Xu}}, \bibinfo {author} {\bibfnamefont {J.}~\bibnamefont
  {Lu}},\ and\ \bibinfo {author} {\bibfnamefont {Y.}~\bibnamefont {Ye}},\
  }\bibfield  {title} {\bibinfo {title} {Odd-even layer-number effect and
  layer-dependent magnetic phase diagrams in
  $\mathrm{MnBi}_{2}\mathrm{Te}_{4}$},\ }\href
  {https://doi.org/10.1103/PhysRevX.11.011003} {\bibfield  {journal} {\bibinfo
  {journal} {Phys. Rev. X}\ }\textbf {\bibinfo {volume} {11}},\ \bibinfo
  {pages} {011003} (\bibinfo {year} {2021})}\BibitemShut {NoStop}%
\bibitem [{\citenamefont {Sun}\ \emph {et~al.}(2019)\citenamefont {Sun},
  \citenamefont {Xia}, \citenamefont {Chen}, \citenamefont {Zhang},
  \citenamefont {Liu}, \citenamefont {Yao}, \citenamefont {Tang}, \citenamefont
  {Zhao}, \citenamefont {Xu},\ and\ \citenamefont
  {Liu}}]{PhysRevLett.123.096401}%
  \BibitemOpen
  \bibfield  {author} {\bibinfo {author} {\bibfnamefont {H.}~\bibnamefont
  {Sun}}, \bibinfo {author} {\bibfnamefont {B.}~\bibnamefont {Xia}}, \bibinfo
  {author} {\bibfnamefont {Z.}~\bibnamefont {Chen}}, \bibinfo {author}
  {\bibfnamefont {Y.}~\bibnamefont {Zhang}}, \bibinfo {author} {\bibfnamefont
  {P.}~\bibnamefont {Liu}}, \bibinfo {author} {\bibfnamefont {Q.}~\bibnamefont
  {Yao}}, \bibinfo {author} {\bibfnamefont {H.}~\bibnamefont {Tang}}, \bibinfo
  {author} {\bibfnamefont {Y.}~\bibnamefont {Zhao}}, \bibinfo {author}
  {\bibfnamefont {H.}~\bibnamefont {Xu}},\ and\ \bibinfo {author}
  {\bibfnamefont {Q.}~\bibnamefont {Liu}},\ }\bibfield  {title} {\bibinfo
  {title} {Rational design principles of the quantum anomalous hall effect in
  superlatticelike magnetic topological insulators},\ }\href
  {https://doi.org/10.1103/PhysRevLett.123.096401} {\bibfield  {journal}
  {\bibinfo  {journal} {Phys. Rev. Lett.}\ }\textbf {\bibinfo {volume} {123}},\
  \bibinfo {pages} {096401} (\bibinfo {year} {2019})}\BibitemShut {NoStop}%
\bibitem [{\citenamefont {Cui}\ \emph {et~al.}(2023)\citenamefont {Cui},
  \citenamefont {Lei}, \citenamefont {Shi}, \citenamefont {Xiang},
  \citenamefont {Wu},\ and\ \citenamefont {Chen}}]{acs.nanolett.2c03773}%
  \BibitemOpen
  \bibfield  {author} {\bibinfo {author} {\bibfnamefont {J.}~\bibnamefont
  {Cui}}, \bibinfo {author} {\bibfnamefont {B.}~\bibnamefont {Lei}}, \bibinfo
  {author} {\bibfnamefont {M.}~\bibnamefont {Shi}}, \bibinfo {author}
  {\bibfnamefont {Z.}~\bibnamefont {Xiang}}, \bibinfo {author} {\bibfnamefont
  {T.}~\bibnamefont {Wu}},\ and\ \bibinfo {author} {\bibfnamefont
  {X.}~\bibnamefont {Chen}},\ }\bibfield  {title} {\bibinfo {title}
  {Layer-dependent magnetic structure and anomalous hall effect in the magnetic
  topological insulator $\mathrm{MnBi}_{2}\mathrm{Te}_{4}$},\ }\href
  {https://doi.org/10.1021/acs.nanolett.2c03773} {\bibfield  {journal}
  {\bibinfo  {journal} {Nano Lett.}\ }\textbf {\bibinfo {volume} {23}},\
  \bibinfo {pages} {1652} (\bibinfo {year} {2023})}\BibitemShut {NoStop}%
\bibitem [{\citenamefont {Zeugner}\ \emph {et~al.}(2019)\citenamefont
  {Zeugner}, \citenamefont {Nietschke}, \citenamefont {Wolter} \emph
  {et~al.}}]{acs.chemmater.8b05017}%
  \BibitemOpen
  \bibfield  {author} {\bibinfo {author} {\bibfnamefont {A.}~\bibnamefont
  {Zeugner}}, \bibinfo {author} {\bibfnamefont {F.}~\bibnamefont {Nietschke}},
  \bibinfo {author} {\bibfnamefont {A.~U.~B.}\ \bibnamefont {Wolter}}, \emph
  {et~al.},\ }\bibfield  {title} {\bibinfo {title} {Chemical aspects of the
  candidate antiferromagnetic topological insulator
  $\mathrm{MnBi}_{2}\mathrm{Te}_{4}$},\ }\href
  {https://doi.org/10.1021/acs.chemmater.8b05017} {\bibfield  {journal}
  {\bibinfo  {journal} {Chem. Mater.}\ }\textbf {\bibinfo {volume} {31}},\
  \bibinfo {pages} {2795} (\bibinfo {year} {2019})}\BibitemShut {NoStop}%
\bibitem [{\citenamefont {L\"upke}\ \emph {et~al.}(2022)\citenamefont
  {L\"upke}, \citenamefont {Pham}, \citenamefont {Zhao}, \citenamefont {Zhou},
  \citenamefont {Lu}, \citenamefont {Briggs}, \citenamefont {Bernholc},
  \citenamefont {Kolmer}, \citenamefont {Teeter}, \citenamefont {Ko},
  \citenamefont {Chang}, \citenamefont {Ganesh},\ and\ \citenamefont
  {Li}}]{PhysRevB.105.035423}%
  \BibitemOpen
  \bibfield  {author} {\bibinfo {author} {\bibfnamefont {F.}~\bibnamefont
  {L\"upke}}, \bibinfo {author} {\bibfnamefont {A.~D.}\ \bibnamefont {Pham}},
  \bibinfo {author} {\bibfnamefont {Y.-F.}\ \bibnamefont {Zhao}}, \bibinfo
  {author} {\bibfnamefont {L.-J.}\ \bibnamefont {Zhou}}, \bibinfo {author}
  {\bibfnamefont {W.}~\bibnamefont {Lu}}, \bibinfo {author} {\bibfnamefont
  {E.}~\bibnamefont {Briggs}}, \bibinfo {author} {\bibfnamefont
  {J.}~\bibnamefont {Bernholc}}, \bibinfo {author} {\bibfnamefont
  {M.}~\bibnamefont {Kolmer}}, \bibinfo {author} {\bibfnamefont
  {J.}~\bibnamefont {Teeter}}, \bibinfo {author} {\bibfnamefont
  {W.}~\bibnamefont {Ko}}, \bibinfo {author} {\bibfnamefont {C.-Z.}\
  \bibnamefont {Chang}}, \bibinfo {author} {\bibfnamefont {P.}~\bibnamefont
  {Ganesh}},\ and\ \bibinfo {author} {\bibfnamefont {A.-P.}\ \bibnamefont
  {Li}},\ }\bibfield  {title} {\bibinfo {title} {Local manifestations of
  thickness-dependent topology and edge states in the topological magnet
  $\mathrm{MnBi}_{2}\mathrm{Te}_{4}$},\ }\href
  {https://doi.org/10.1103/PhysRevB.105.035423} {\bibfield  {journal} {\bibinfo
   {journal} {Phys. Rev. B}\ }\textbf {\bibinfo {volume} {105}},\ \bibinfo
  {pages} {035423} (\bibinfo {year} {2022})}\BibitemShut {NoStop}%
\bibitem [{\citenamefont {Mong}\ \emph {et~al.}(2010)\citenamefont {Mong},
  \citenamefont {Essin},\ and\ \citenamefont {Moore}}]{PhysRevB.81.245209}%
  \BibitemOpen
  \bibfield  {author} {\bibinfo {author} {\bibfnamefont {R.~S.~K.}\
  \bibnamefont {Mong}}, \bibinfo {author} {\bibfnamefont {A.~M.}\ \bibnamefont
  {Essin}},\ and\ \bibinfo {author} {\bibfnamefont {J.~E.}\ \bibnamefont
  {Moore}},\ }\bibfield  {title} {\bibinfo {title} {Antiferromagnetic
  topological insulators},\ }\href {https://doi.org/10.1103/PhysRevB.81.245209}
  {\bibfield  {journal} {\bibinfo  {journal} {Phys. Rev. B}\ }\textbf {\bibinfo
  {volume} {81}},\ \bibinfo {pages} {245209} (\bibinfo {year}
  {2010})}\BibitemShut {NoStop}%
\bibitem [{\citenamefont {Tang}\ \emph {et~al.}(2016)\citenamefont {Tang},
  \citenamefont {Zhou}, \citenamefont {Xu},\ and\ \citenamefont
  {Zhang}}]{2016Dirac}%
  \BibitemOpen
  \bibfield  {author} {\bibinfo {author} {\bibfnamefont {P.}~\bibnamefont
  {Tang}}, \bibinfo {author} {\bibfnamefont {Q.}~\bibnamefont {Zhou}}, \bibinfo
  {author} {\bibfnamefont {G.}~\bibnamefont {Xu}},\ and\ \bibinfo {author}
  {\bibfnamefont {S.-C.}\ \bibnamefont {Zhang}},\ }\bibfield  {title} {\bibinfo
  {title} {Dirac fermions in an antiferromagnetic semimetal},\ }\href
  {https://doi.org/10.1038/NPHYS3839} {\bibfield  {journal} {\bibinfo
  {journal} {Nat. Phys.}\ }\textbf {\bibinfo {volume} {12}},\ \bibinfo {pages}
  {1100} (\bibinfo {year} {2016})}\BibitemShut {NoStop}%
\bibitem [{\citenamefont {\ifmmode~\check{S}\else \v{S}\fi{}mejkal}\ \emph
  {et~al.}(2017)\citenamefont {\ifmmode~\check{S}\else \v{S}\fi{}mejkal},
  \citenamefont {\ifmmode~\check{Z}\else \v{Z}\fi{}elezn\'y}, \citenamefont
  {Sinova},\ and\ \citenamefont {Jungwirth}}]{PhysRevLett.118.106402}%
  \BibitemOpen
  \bibfield  {author} {\bibinfo {author} {\bibfnamefont {L.}~\bibnamefont
  {\ifmmode~\check{S}\else \v{S}\fi{}mejkal}}, \bibinfo {author} {\bibfnamefont
  {J.}~\bibnamefont {\ifmmode~\check{Z}\else \v{Z}\fi{}elezn\'y}}, \bibinfo
  {author} {\bibfnamefont {J.}~\bibnamefont {Sinova}},\ and\ \bibinfo {author}
  {\bibfnamefont {T.}~\bibnamefont {Jungwirth}},\ }\bibfield  {title} {\bibinfo
  {title} {Electric control of dirac quasiparticles by spin-orbit torque in an
  antiferromagnet},\ }\href {https://doi.org/10.1103/PhysRevLett.118.106402}
  {\bibfield  {journal} {\bibinfo  {journal} {Phys. Rev. Lett.}\ }\textbf
  {\bibinfo {volume} {118}},\ \bibinfo {pages} {106402} (\bibinfo {year}
  {2017})}\BibitemShut {NoStop}%
\bibitem [{\citenamefont {Liu}\ \emph {et~al.}(2019{\natexlab{b}})\citenamefont
  {Liu}, \citenamefont {Meng},\ and\ \citenamefont {Sun}}]{nanolett.9b00948}%
  \BibitemOpen
  \bibfield  {author} {\bibinfo {author} {\bibfnamefont {J.}~\bibnamefont
  {Liu}}, \bibinfo {author} {\bibfnamefont {S.}~\bibnamefont {Meng}},\ and\
  \bibinfo {author} {\bibfnamefont {J.-T.}\ \bibnamefont {Sun}},\ }\bibfield
  {title} {\bibinfo {title} {Spin-orientation-dependent topological states in
  two-dimensional antiferromagnetic $\mathrm{NiTl}_{2}\mathrm{S}_{4}$
  monolayers},\ }\href {https://doi.org/10.1021/acs.nanolett.9b00948}
  {\bibfield  {journal} {\bibinfo  {journal} {Nano Lett.}\ }\textbf {\bibinfo
  {volume} {19}},\ \bibinfo {pages} {3321} (\bibinfo {year}
  {2019}{\natexlab{b}})}\BibitemShut {NoStop}%
\bibitem [{\citenamefont {Shao}\ \emph {et~al.}(2019)\citenamefont {Shao},
  \citenamefont {Gurung}, \citenamefont {Zhang},\ and\ \citenamefont
  {Tsymbal}}]{PhysRevLett.122.077203}%
  \BibitemOpen
  \bibfield  {author} {\bibinfo {author} {\bibfnamefont {D.-F.}\ \bibnamefont
  {Shao}}, \bibinfo {author} {\bibfnamefont {G.}~\bibnamefont {Gurung}},
  \bibinfo {author} {\bibfnamefont {S.-H.}\ \bibnamefont {Zhang}},\ and\
  \bibinfo {author} {\bibfnamefont {E.~Y.}\ \bibnamefont {Tsymbal}},\
  }\bibfield  {title} {\bibinfo {title} {Dirac nodal line metal for topological
  antiferromagnetic spintronics},\ }\href
  {https://doi.org/10.1103/PhysRevLett.122.077203} {\bibfield  {journal}
  {\bibinfo  {journal} {Phys. Rev. Lett.}\ }\textbf {\bibinfo {volume} {122}},\
  \bibinfo {pages} {077203} (\bibinfo {year} {2019})}\BibitemShut {NoStop}%
\bibitem [{\citenamefont {Niu}\ \emph {et~al.}(2020)\citenamefont {Niu},
  \citenamefont {Wang}, \citenamefont {Mao}, \citenamefont {Huang},
  \citenamefont {Mokrousov},\ and\ \citenamefont
  {Dai}}]{PhysRevLett.124.066401}%
  \BibitemOpen
  \bibfield  {author} {\bibinfo {author} {\bibfnamefont {C.}~\bibnamefont
  {Niu}}, \bibinfo {author} {\bibfnamefont {H.}~\bibnamefont {Wang}}, \bibinfo
  {author} {\bibfnamefont {N.}~\bibnamefont {Mao}}, \bibinfo {author}
  {\bibfnamefont {B.}~\bibnamefont {Huang}}, \bibinfo {author} {\bibfnamefont
  {Y.}~\bibnamefont {Mokrousov}},\ and\ \bibinfo {author} {\bibfnamefont
  {Y.}~\bibnamefont {Dai}},\ }\bibfield  {title} {\bibinfo {title}
  {Antiferromagnetic topological insulator with nonsymmorphic protection in two
  dimensions},\ }\href {https://doi.org/10.1103/PhysRevLett.124.066401}
  {\bibfield  {journal} {\bibinfo  {journal} {Phys. Rev. Lett.}\ }\textbf
  {\bibinfo {volume} {124}},\ \bibinfo {pages} {066401} (\bibinfo {year}
  {2020})}\BibitemShut {NoStop}%
\bibitem [{\citenamefont {Smejkal}\ \emph {et~al.}(2018)\citenamefont
  {Smejkal}, \citenamefont {Mokrousov}, \citenamefont {Yan},\ and\
  \citenamefont {MacDonald}}]{2019Topoantifer}%
  \BibitemOpen
  \bibfield  {author} {\bibinfo {author} {\bibfnamefont {L.}~\bibnamefont
  {Smejkal}}, \bibinfo {author} {\bibfnamefont {Y.}~\bibnamefont {Mokrousov}},
  \bibinfo {author} {\bibfnamefont {B.}~\bibnamefont {Yan}},\ and\ \bibinfo
  {author} {\bibfnamefont {A.~H.}\ \bibnamefont {MacDonald}},\ }\bibfield
  {title} {\bibinfo {title} {Topological antiferromagnetic spintronics},\
  }\href {https://doi.org/10.1038/s41567-018-0064-5} {\bibfield  {journal}
  {\bibinfo  {journal} {Nat. Phys.}\ }\textbf {\bibinfo {volume} {14}},\
  \bibinfo {pages} {242} (\bibinfo {year} {2018})}\BibitemShut {NoStop}%
\bibitem [{\citenamefont {Otrokov}\ \emph
  {et~al.}(2019{\natexlab{b}})\citenamefont {Otrokov}, \citenamefont
  {Klimovskikh}, \citenamefont {Bentmann} \emph {et~al.}}]{2019Prediction}%
  \BibitemOpen
  \bibfield  {author} {\bibinfo {author} {\bibfnamefont {M.~M.}\ \bibnamefont
  {Otrokov}}, \bibinfo {author} {\bibfnamefont {I.~I.}\ \bibnamefont
  {Klimovskikh}}, \bibinfo {author} {\bibfnamefont {H.}~\bibnamefont
  {Bentmann}}, \emph {et~al.},\ }\bibfield  {title} {\bibinfo {title}
  {Prediction and observation of an antiferromagnetic topological insulator},\
  }\href {https://doi.org/10.1038/s41586-019-1840-9} {\bibfield  {journal}
  {\bibinfo  {journal} {Nature}\ }\textbf {\bibinfo {volume} {576}},\ \bibinfo
  {pages} {416} (\bibinfo {year} {2019}{\natexlab{b}})}\BibitemShut {NoStop}%
\bibitem [{\citenamefont {Wu}\ \emph {et~al.}(2019)\citenamefont {Wu},
  \citenamefont {Liu}, \citenamefont {Sasase}, \citenamefont {Ienaga},
  \citenamefont {Obata}, \citenamefont {Yukawa}, \citenamefont {Horiba},
  \citenamefont {Kumigashira}, \citenamefont {Okuma}, \citenamefont
  {Inoshita},\ and\ \citenamefont {Hosono}}]{sciadv.aax9989}%
  \BibitemOpen
  \bibfield  {author} {\bibinfo {author} {\bibfnamefont {J.}~\bibnamefont
  {Wu}}, \bibinfo {author} {\bibfnamefont {F.}~\bibnamefont {Liu}}, \bibinfo
  {author} {\bibfnamefont {M.}~\bibnamefont {Sasase}}, \bibinfo {author}
  {\bibfnamefont {K.}~\bibnamefont {Ienaga}}, \bibinfo {author} {\bibfnamefont
  {Y.}~\bibnamefont {Obata}}, \bibinfo {author} {\bibfnamefont
  {R.}~\bibnamefont {Yukawa}}, \bibinfo {author} {\bibfnamefont
  {K.}~\bibnamefont {Horiba}}, \bibinfo {author} {\bibfnamefont
  {H.}~\bibnamefont {Kumigashira}}, \bibinfo {author} {\bibfnamefont
  {S.}~\bibnamefont {Okuma}}, \bibinfo {author} {\bibfnamefont
  {T.}~\bibnamefont {Inoshita}},\ and\ \bibinfo {author} {\bibfnamefont
  {H.}~\bibnamefont {Hosono}},\ }\bibfield  {title} {\bibinfo {title} {Natural
  van der waals heterostructural single crystals with both magnetic and
  topological properties},\ }\href {https://doi.org/10.1126/sciadv.aax9989}
  {\bibfield  {journal} {\bibinfo  {journal} {Sci. Adv.}\ }\textbf {\bibinfo
  {volume} {5}},\ \bibinfo {pages} {eaax9989} (\bibinfo {year}
  {2019})}\BibitemShut {NoStop}%
\bibitem [{\citenamefont {Shi}\ \emph {et~al.}(2019)\citenamefont {Shi},
  \citenamefont {Lei}, \citenamefont {Zhu}, \citenamefont {Ma}, \citenamefont
  {Cui}, \citenamefont {Sun}, \citenamefont {Ying},\ and\ \citenamefont
  {Chen}}]{PhysRevB.100.155144}%
  \BibitemOpen
  \bibfield  {author} {\bibinfo {author} {\bibfnamefont {M.~Z.}\ \bibnamefont
  {Shi}}, \bibinfo {author} {\bibfnamefont {B.}~\bibnamefont {Lei}}, \bibinfo
  {author} {\bibfnamefont {C.~S.}\ \bibnamefont {Zhu}}, \bibinfo {author}
  {\bibfnamefont {D.~H.}\ \bibnamefont {Ma}}, \bibinfo {author} {\bibfnamefont
  {J.~H.}\ \bibnamefont {Cui}}, \bibinfo {author} {\bibfnamefont {Z.~L.}\
  \bibnamefont {Sun}}, \bibinfo {author} {\bibfnamefont {J.~J.}\ \bibnamefont
  {Ying}},\ and\ \bibinfo {author} {\bibfnamefont {X.~H.}\ \bibnamefont
  {Chen}},\ }\bibfield  {title} {\bibinfo {title} {Magnetic and transport
  properties in the magnetic topological insulators
  $\mathrm{MnB}{\mathrm{i}}_{2}\mathrm{T}{\mathrm{e}}_{4}{(\mathrm{B}{\mathrm{i}}_{2}\mathrm{T}{\mathrm{e}}_{3})}_{n}$
  ($n=1,2$)},\ }\href {https://doi.org/10.1103/PhysRevB.100.155144} {\bibfield
  {journal} {\bibinfo  {journal} {Phys. Rev. B}\ }\textbf {\bibinfo {volume}
  {100}},\ \bibinfo {pages} {155144} (\bibinfo {year} {2019})}\BibitemShut
  {NoStop}%
\bibitem [{\citenamefont {Vidal}\ \emph {et~al.}(2019)\citenamefont {Vidal},
  \citenamefont {Zeugner}, \citenamefont {Facio} \emph
  {et~al.}}]{PhysRevX.9.041065}%
  \BibitemOpen
  \bibfield  {author} {\bibinfo {author} {\bibfnamefont {R.~C.}\ \bibnamefont
  {Vidal}}, \bibinfo {author} {\bibfnamefont {A.}~\bibnamefont {Zeugner}},
  \bibinfo {author} {\bibfnamefont {J.~I.}\ \bibnamefont {Facio}}, \emph
  {et~al.},\ }\bibfield  {title} {\bibinfo {title} {Topological electronic
  structure and intrinsic magnetization in $\mathrm{MnBi}_{4}\mathrm{Te}_{7}$:
  A $\mathrm{Bi}_{2}\mathrm{Te}_{3}$ derivative with a periodic mn
  sublattice},\ }\href {https://doi.org/10.1103/PhysRevX.9.041065} {\bibfield
  {journal} {\bibinfo  {journal} {Phys. Rev. X}\ }\textbf {\bibinfo {volume}
  {9}},\ \bibinfo {pages} {041065} (\bibinfo {year} {2019})}\BibitemShut
  {NoStop}%
\bibitem [{\citenamefont {Hu}\ \emph {et~al.}(2020{\natexlab{a}})\citenamefont
  {Hu}, \citenamefont {Gordon}, \citenamefont {Liu}, \citenamefont {Liu},
  \citenamefont {Zhou}, \citenamefont {Hao}, \citenamefont {Narayan},
  \citenamefont {Emmanouilidou}, \citenamefont {Sun}, \citenamefont {Liu},
  \citenamefont {Brawer}, \citenamefont {Ramirez}, \citenamefont {Ding},
  \citenamefont {Cao}, \citenamefont {Liu}, \citenamefont {Dessau},\ and\
  \citenamefont {Ni}}]{vanderWaals13814}%
  \BibitemOpen
  \bibfield  {author} {\bibinfo {author} {\bibfnamefont {C.}~\bibnamefont
  {Hu}}, \bibinfo {author} {\bibfnamefont {K.~N.}\ \bibnamefont {Gordon}},
  \bibinfo {author} {\bibfnamefont {P.}~\bibnamefont {Liu}}, \bibinfo {author}
  {\bibfnamefont {J.}~\bibnamefont {Liu}}, \bibinfo {author} {\bibfnamefont
  {X.}~\bibnamefont {Zhou}}, \bibinfo {author} {\bibfnamefont {P.}~\bibnamefont
  {Hao}}, \bibinfo {author} {\bibfnamefont {D.}~\bibnamefont {Narayan}},
  \bibinfo {author} {\bibfnamefont {E.}~\bibnamefont {Emmanouilidou}}, \bibinfo
  {author} {\bibfnamefont {H.}~\bibnamefont {Sun}}, \bibinfo {author}
  {\bibfnamefont {Y.}~\bibnamefont {Liu}}, \bibinfo {author} {\bibfnamefont
  {H.}~\bibnamefont {Brawer}}, \bibinfo {author} {\bibfnamefont {A.~P.}\
  \bibnamefont {Ramirez}}, \bibinfo {author} {\bibfnamefont {L.}~\bibnamefont
  {Ding}}, \bibinfo {author} {\bibfnamefont {H.}~\bibnamefont {Cao}}, \bibinfo
  {author} {\bibfnamefont {Q.}~\bibnamefont {Liu}}, \bibinfo {author}
  {\bibfnamefont {D.}~\bibnamefont {Dessau}},\ and\ \bibinfo {author}
  {\bibfnamefont {N.}~\bibnamefont {Ni}},\ }\bibfield  {title} {\bibinfo
  {title} {A van der waals antiferromagnetic topological insulator with weak
  interlayer magnetic coupling},\ }\href
  {https://doi.org/10.1038/s41467-019-13814-x} {\bibfield  {journal} {\bibinfo
  {journal} {Nat. Commun.}\ }\textbf {\bibinfo {volume} {11}},\ \bibinfo
  {pages} {97} (\bibinfo {year} {2020}{\natexlab{a}})}\BibitemShut {NoStop}%
\bibitem [{\citenamefont {Klimovskikh}\ \emph {et~al.}(2020)\citenamefont
  {Klimovskikh}, \citenamefont {Otrokov}, \citenamefont {Estyunin} \emph
  {et~al.}}]{npjTunable}%
  \BibitemOpen
  \bibfield  {author} {\bibinfo {author} {\bibfnamefont {I.~I.}\ \bibnamefont
  {Klimovskikh}}, \bibinfo {author} {\bibfnamefont {M.~M.}\ \bibnamefont
  {Otrokov}}, \bibinfo {author} {\bibfnamefont {D.}~\bibnamefont {Estyunin}},
  \emph {et~al.},\ }\bibfield  {title} {\bibinfo {title} {Tunable {3D/2D}
  magnetism in the
  ($\mathrm{MnBi}_{2}\mathrm{Te}_{4}$)($\mathrm{Bi}_{2}\mathrm{Te}_{4})_{m}$
  topological insulators family},\ }\href
  {https://doi.org/10.1038/s41535-020-00255-9} {\bibfield  {journal} {\bibinfo
  {journal} {npj Quantum Mater.}\ }\textbf {\bibinfo {volume} {5}},\ \bibinfo
  {pages} {54} (\bibinfo {year} {2020})}\BibitemShut {NoStop}%
\bibitem [{\citenamefont {Hu}\ \emph {et~al.}(2020{\natexlab{b}})\citenamefont
  {Hu}, \citenamefont {Xu}, \citenamefont {Shi}, \citenamefont {Luo},
  \citenamefont {Peng}, \citenamefont {Wang}, \citenamefont {Ying},
  \citenamefont {Wu}, \citenamefont {Liu}, \citenamefont {Zhang}, \citenamefont
  {Chen}, \citenamefont {Xu}, \citenamefont {Chen},\ and\ \citenamefont
  {He}}]{PhysRevB.101.161113}%
  \BibitemOpen
  \bibfield  {author} {\bibinfo {author} {\bibfnamefont {Y.}~\bibnamefont
  {Hu}}, \bibinfo {author} {\bibfnamefont {L.}~\bibnamefont {Xu}}, \bibinfo
  {author} {\bibfnamefont {M.}~\bibnamefont {Shi}}, \bibinfo {author}
  {\bibfnamefont {A.}~\bibnamefont {Luo}}, \bibinfo {author} {\bibfnamefont
  {S.}~\bibnamefont {Peng}}, \bibinfo {author} {\bibfnamefont {Z.~Y.}\
  \bibnamefont {Wang}}, \bibinfo {author} {\bibfnamefont {J.~J.}\ \bibnamefont
  {Ying}}, \bibinfo {author} {\bibfnamefont {T.}~\bibnamefont {Wu}}, \bibinfo
  {author} {\bibfnamefont {Z.~K.}\ \bibnamefont {Liu}}, \bibinfo {author}
  {\bibfnamefont {C.~F.}\ \bibnamefont {Zhang}}, \bibinfo {author}
  {\bibfnamefont {Y.~L.}\ \bibnamefont {Chen}}, \bibinfo {author}
  {\bibfnamefont {G.}~\bibnamefont {Xu}}, \bibinfo {author} {\bibfnamefont
  {X.-H.}\ \bibnamefont {Chen}},\ and\ \bibinfo {author} {\bibfnamefont
  {J.-F.}\ \bibnamefont {He}},\ }\bibfield  {title} {\bibinfo {title}
  {Universal gapless dirac cone and tunable topological states in
  ${(\mathrm{MnB}{\mathrm{i}}_{2}\mathrm{T}{\mathrm{e}}_{4})}_{m}{(\mathrm{B}{\mathrm{i}}_{2}\mathrm{T}{\mathrm{e}}_{3})}_{n}$
  heterostructures},\ }\href {https://doi.org/10.1103/PhysRevB.101.161113}
  {\bibfield  {journal} {\bibinfo  {journal} {Phys. Rev. B}\ }\textbf {\bibinfo
  {volume} {101}},\ \bibinfo {pages} {161113} (\bibinfo {year}
  {2020}{\natexlab{b}})}\BibitemShut {NoStop}%
\bibitem [{\citenamefont {Yan}\ \emph {et~al.}(2020)\citenamefont {Yan},
  \citenamefont {Liu}, \citenamefont {Parker}, \citenamefont {Wu},
  \citenamefont {Aczel}, \citenamefont {Matsuda}, \citenamefont {McGuire},\
  and\ \citenamefont {Sales}}]{PhysRevMaterials.4.054202}%
  \BibitemOpen
  \bibfield  {author} {\bibinfo {author} {\bibfnamefont {J.-Q.}\ \bibnamefont
  {Yan}}, \bibinfo {author} {\bibfnamefont {Y.~H.}\ \bibnamefont {Liu}},
  \bibinfo {author} {\bibfnamefont {D.~S.}\ \bibnamefont {Parker}}, \bibinfo
  {author} {\bibfnamefont {Y.}~\bibnamefont {Wu}}, \bibinfo {author}
  {\bibfnamefont {A.~A.}\ \bibnamefont {Aczel}}, \bibinfo {author}
  {\bibfnamefont {M.}~\bibnamefont {Matsuda}}, \bibinfo {author} {\bibfnamefont
  {M.~A.}\ \bibnamefont {McGuire}},\ and\ \bibinfo {author} {\bibfnamefont
  {B.~C.}\ \bibnamefont {Sales}},\ }\bibfield  {title} {\bibinfo {title}
  {A-type antiferromagnetic order in $\mathrm{MnBi}_{4}\mathrm{Te}_{7}$ and
  $\mathrm{MnBi}_{6}\mathrm{Te}_{10}$ single crystals},\ }\href
  {https://doi.org/10.1103/PhysRevMaterials.4.054202} {\bibfield  {journal}
  {\bibinfo  {journal} {Phys. Rev. Mater.}\ }\textbf {\bibinfo {volume} {4}},\
  \bibinfo {pages} {054202} (\bibinfo {year} {2020})}\BibitemShut {NoStop}%
\bibitem [{\citenamefont {Mu}\ \emph {et~al.}(2022)\citenamefont {Mu},
  \citenamefont {Zhao}, \citenamefont {Zhang},\ and\ \citenamefont
  {Wang}}]{22NPJmasskink}%
  \BibitemOpen
  \bibfield  {author} {\bibinfo {author} {\bibfnamefont {H.}~\bibnamefont
  {Mu}}, \bibinfo {author} {\bibfnamefont {G.}~\bibnamefont {Zhao}}, \bibinfo
  {author} {\bibfnamefont {H.}~\bibnamefont {Zhang}},\ and\ \bibinfo {author}
  {\bibfnamefont {Z.}~\bibnamefont {Wang}},\ }\bibfield  {title} {\bibinfo
  {title} {Antiferromagnetic second-order topological insulator with fractional
  mass-kink},\ }\href {https://doi.org/10.1038/s41524-022-00761-3} {\bibfield
  {journal} {\bibinfo  {journal} {npj Comput. Mater.}\ }\textbf {\bibinfo
  {volume} {8}},\ \bibinfo {pages} {82} (\bibinfo {year} {2022})}\BibitemShut
  {NoStop}%
\bibitem [{\citenamefont {Luo}\ \emph {et~al.}(2022)\citenamefont {Luo},
  \citenamefont {Song},\ and\ \citenamefont {Xu}}]{22Fragile}%
  \BibitemOpen
  \bibfield  {author} {\bibinfo {author} {\bibfnamefont {A.}~\bibnamefont
  {Luo}}, \bibinfo {author} {\bibfnamefont {Z.}~\bibnamefont {Song}},\ and\
  \bibinfo {author} {\bibfnamefont {G.}~\bibnamefont {Xu}},\ }\bibfield
  {title} {\bibinfo {title} {Fragile topological band in the checkerboard
  antiferromagnetic monolayer {FeSe}},\ }\href
  {https://doi.org/10.1038/s41524-022-00707-9} {\bibfield  {journal} {\bibinfo
  {journal} {npj Comput. Mater.}\ }\textbf {\bibinfo {volume} {8}},\ \bibinfo
  {pages} {26} (\bibinfo {year} {2022})}\BibitemShut {NoStop}%
\bibitem [{\citenamefont {Zhang}\ \emph {et~al.}(2022)\citenamefont {Zhang},
  \citenamefont {Yu}, \citenamefont {Liu},\ and\ \citenamefont
  {Yao}}]{ZHANG2022108153}%
  \BibitemOpen
  \bibfield  {author} {\bibinfo {author} {\bibfnamefont {Z.}~\bibnamefont
  {Zhang}}, \bibinfo {author} {\bibfnamefont {Z.-M.}\ \bibnamefont {Yu}},
  \bibinfo {author} {\bibfnamefont {G.-B.}\ \bibnamefont {Liu}},\ and\ \bibinfo
  {author} {\bibfnamefont {Y.}~\bibnamefont {Yao}},\ }\bibfield  {title}
  {\bibinfo {title} {Magnetictb: A package for tight-binding model of magnetic
  and non-magnetic materials},\ }\href
  {https://doi.org/https://doi.org/10.1016/j.cpc.2021.108153} {\bibfield
  {journal} {\bibinfo  {journal} {Comput. Phys. Commun.}\ }\textbf {\bibinfo
  {volume} {270}},\ \bibinfo {pages} {108153} (\bibinfo {year}
  {2022})}\BibitemShut {NoStop}%
\bibitem [{\citenamefont {Gmitra}\ \emph {et~al.}(2013)\citenamefont {Gmitra},
  \citenamefont {Kochan},\ and\ \citenamefont
  {Fabian}}]{PhysRevLett.110.246602}%
  \BibitemOpen
  \bibfield  {author} {\bibinfo {author} {\bibfnamefont {M.}~\bibnamefont
  {Gmitra}}, \bibinfo {author} {\bibfnamefont {D.}~\bibnamefont {Kochan}},\
  and\ \bibinfo {author} {\bibfnamefont {J.}~\bibnamefont {Fabian}},\
  }\bibfield  {title} {\bibinfo {title} {Spin-orbit coupling in hydrogenated
  graphene},\ }\href {https://doi.org/10.1103/PhysRevLett.110.246602}
  {\bibfield  {journal} {\bibinfo  {journal} {Phys. Rev. Lett.}\ }\textbf
  {\bibinfo {volume} {110}},\ \bibinfo {pages} {246602} (\bibinfo {year}
  {2013})}\BibitemShut {NoStop}%
\bibitem [{\citenamefont {Irmer}\ \emph {et~al.}(2015)\citenamefont {Irmer},
  \citenamefont {Frank}, \citenamefont {Putz}, \citenamefont {Gmitra},
  \citenamefont {Kochan},\ and\ \citenamefont {Fabian}}]{PhysRevB.91.115141}%
  \BibitemOpen
  \bibfield  {author} {\bibinfo {author} {\bibfnamefont {S.}~\bibnamefont
  {Irmer}}, \bibinfo {author} {\bibfnamefont {T.}~\bibnamefont {Frank}},
  \bibinfo {author} {\bibfnamefont {S.}~\bibnamefont {Putz}}, \bibinfo {author}
  {\bibfnamefont {M.}~\bibnamefont {Gmitra}}, \bibinfo {author} {\bibfnamefont
  {D.}~\bibnamefont {Kochan}},\ and\ \bibinfo {author} {\bibfnamefont
  {J.}~\bibnamefont {Fabian}},\ }\bibfield  {title} {\bibinfo {title}
  {Spin-orbit coupling in fluorinated graphene},\ }\href
  {https://doi.org/10.1103/PhysRevB.91.115141} {\bibfield  {journal} {\bibinfo
  {journal} {Phys. Rev. B}\ }\textbf {\bibinfo {volume} {91}},\ \bibinfo
  {pages} {115141} (\bibinfo {year} {2015})}\BibitemShut {NoStop}%
\bibitem [{SM()}]{SM}%
  \BibitemOpen
  \href@noop {} {\bibinfo  {journal} {{See Supplemental Material for detailed
  computational methods of first-principles calculations, the minimum lattice
  model of the AFM MnBi$_{2}$Te$_{4}$ bilayer, and rotational topological
  invariants and corner charge for AFM MnBi$_2$Te$_4$ bilayer and Bi$_2$Te$_3$
  QL systems, which inlcude Refs.
  ~\cite{Xu2020,Elcoro2021,PhysRevB.103.205123,PhysRevLett.122.107202,SMPhysRev.136.B864,SMPhysRev.140.A1133,SMPhysRevB.54.11169,SMPhysRevLett.77.3865,SMMostofi2014,SMSancho_1985,SMWU2017,SMDensity,ZHANG2022108153}}}\
  }\BibitemShut {NoStop}%
\bibitem [{\citenamefont {Hohenberg}\ and\ \citenamefont
  {Kohn}(1964)}]{SMPhysRev.136.B864}%
  \BibitemOpen
\bibfield  {journal} {  }\bibfield  {author} {\bibinfo {author} {\bibfnamefont
  {P.}~\bibnamefont {Hohenberg}}\ and\ \bibinfo {author} {\bibfnamefont
  {W.}~\bibnamefont {Kohn}},\ }\bibfield  {title} {\bibinfo {title}
  {Inhomogeneous electron gas},\ }\href
  {https://doi.org/10.1103/PhysRev.136.B864} {\bibfield  {journal} {\bibinfo
  {journal} {Phys. Rev.}\ }\textbf {\bibinfo {volume} {136}},\ \bibinfo {pages}
  {B864} (\bibinfo {year} {1964})}\BibitemShut {NoStop}%
\bibitem [{\citenamefont {Kohn}\ and\ \citenamefont
  {Sham}(1965)}]{SMPhysRev.140.A1133}%
  \BibitemOpen
  \bibfield  {author} {\bibinfo {author} {\bibfnamefont {W.}~\bibnamefont
  {Kohn}}\ and\ \bibinfo {author} {\bibfnamefont {L.~J.}\ \bibnamefont
  {Sham}},\ }\bibfield  {title} {\bibinfo {title} {Self-consistent equations
  including exchange and correlation effects},\ }\href
  {https://doi.org/10.1103/PhysRev.140.A1133} {\bibfield  {journal} {\bibinfo
  {journal} {Phys. Rev.}\ }\textbf {\bibinfo {volume} {140}},\ \bibinfo {pages}
  {A1133} (\bibinfo {year} {1965})}\BibitemShut {NoStop}%
\bibitem [{\citenamefont {Sancho}\ \emph {et~al.}(1985)\citenamefont {Sancho},
  \citenamefont {Sancho}, \citenamefont {Sancho},\ and\ \citenamefont
  {Rubio}}]{SMSancho_1985}%
  \BibitemOpen
  \bibfield  {author} {\bibinfo {author} {\bibfnamefont {M.~P.~L.}\
  \bibnamefont {Sancho}}, \bibinfo {author} {\bibfnamefont {J.~M.~L.}\
  \bibnamefont {Sancho}}, \bibinfo {author} {\bibfnamefont {J.~M.~L.}\
  \bibnamefont {Sancho}},\ and\ \bibinfo {author} {\bibfnamefont
  {J.}~\bibnamefont {Rubio}},\ }\bibfield  {title} {\bibinfo {title} {Highly
  convergent schemes for the calculation of bulk and surface green functions},\
  }\href {https://doi.org/10.1088/0305-4608/15/4/009} {\bibfield  {journal}
  {\bibinfo  {journal} {J. Phys. F: Met. Phys}\ }\textbf {\bibinfo {volume}
  {15}},\ \bibinfo {pages} {851} (\bibinfo {year} {1985})}\BibitemShut
  {NoStop}%
\bibitem [{\citenamefont {Kresse}\ and\ \citenamefont
  {Furthm\"uller}(1996)}]{SMPhysRevB.54.11169}%
  \BibitemOpen
  \bibfield  {author} {\bibinfo {author} {\bibfnamefont {G.}~\bibnamefont
  {Kresse}}\ and\ \bibinfo {author} {\bibfnamefont {J.}~\bibnamefont
  {Furthm\"uller}},\ }\bibfield  {title} {\bibinfo {title} {Efficient iterative
  schemes for ab initio total-energy calculations using a plane-wave basis
  set},\ }\href {https://doi.org/10.1103/PhysRevB.54.11169} {\bibfield
  {journal} {\bibinfo  {journal} {Phys. Rev. B}\ }\textbf {\bibinfo {volume}
  {54}},\ \bibinfo {pages} {11169} (\bibinfo {year} {1996})}\BibitemShut
  {NoStop}%
\bibitem [{\citenamefont {Perdew}\ \emph {et~al.}(1996)\citenamefont {Perdew},
  \citenamefont {Burke},\ and\ \citenamefont
  {Ernzerhof}}]{SMPhysRevLett.77.3865}%
  \BibitemOpen
  \bibfield  {author} {\bibinfo {author} {\bibfnamefont {J.~P.}\ \bibnamefont
  {Perdew}}, \bibinfo {author} {\bibfnamefont {K.}~\bibnamefont {Burke}},\ and\
  \bibinfo {author} {\bibfnamefont {M.}~\bibnamefont {Ernzerhof}},\ }\bibfield
  {title} {\bibinfo {title} {Generalized gradient approximation made simple},\
  }\href {https://doi.org/10.1103/PhysRevLett.77.3865} {\bibfield  {journal}
  {\bibinfo  {journal} {Phys. Rev. Lett.}\ }\textbf {\bibinfo {volume} {77}},\
  \bibinfo {pages} {3865} (\bibinfo {year} {1996})}\BibitemShut {NoStop}%
\bibitem [{\citenamefont {Mostofi}\ \emph {et~al.}(2014)\citenamefont
  {Mostofi}, \citenamefont {Yates}, \citenamefont {Pizzi}, \citenamefont {Lee},
  \citenamefont {Souza}, \citenamefont {Vanderbilt},\ and\ \citenamefont
  {Marzari}}]{SMMostofi2014}%
  \BibitemOpen
  \bibfield  {author} {\bibinfo {author} {\bibfnamefont {A.~A.}\ \bibnamefont
  {Mostofi}}, \bibinfo {author} {\bibfnamefont {J.~R.}\ \bibnamefont {Yates}},
  \bibinfo {author} {\bibfnamefont {G.}~\bibnamefont {Pizzi}}, \bibinfo
  {author} {\bibfnamefont {Y.-S.}\ \bibnamefont {Lee}}, \bibinfo {author}
  {\bibfnamefont {I.}~\bibnamefont {Souza}}, \bibinfo {author} {\bibfnamefont
  {D.}~\bibnamefont {Vanderbilt}},\ and\ \bibinfo {author} {\bibfnamefont
  {N.}~\bibnamefont {Marzari}},\ }\bibfield  {title} {\bibinfo {title} {An
  updated version of wannier90: A tool for obtaining maximally-localised
  wannier functions},\ }\href
  {https://doi.org/https://doi.org/10.1016/j.cpc.2014.05.003} {\bibfield
  {journal} {\bibinfo  {journal} {Comput. Phys. Commun.}\ }\textbf {\bibinfo
  {volume} {185}},\ \bibinfo {pages} {2309} (\bibinfo {year}
  {2014})}\BibitemShut {NoStop}%
\bibitem [{\citenamefont {Wu}\ \emph {et~al.}(2018)\citenamefont {Wu},
  \citenamefont {Zhang}, \citenamefont {Song}, \citenamefont {Troyer},\ and\
  \citenamefont {Soluyanov}}]{SMWU2017}%
  \BibitemOpen
  \bibfield  {author} {\bibinfo {author} {\bibfnamefont {Q.}~\bibnamefont
  {Wu}}, \bibinfo {author} {\bibfnamefont {S.}~\bibnamefont {Zhang}}, \bibinfo
  {author} {\bibfnamefont {H.-F.}\ \bibnamefont {Song}}, \bibinfo {author}
  {\bibfnamefont {M.}~\bibnamefont {Troyer}},\ and\ \bibinfo {author}
  {\bibfnamefont {A.~A.}\ \bibnamefont {Soluyanov}},\ }\bibfield  {title}
  {\bibinfo {title} {Wanniertools : An open-source software package for novel
  topological materials},\ }\href {https://doi.org/10.1016/j.cpc.2017.09.033}
  {\bibfield  {journal} {\bibinfo  {journal} {Comput. Phys. Commun.}\ }\textbf
  {\bibinfo {volume} {224}},\ \bibinfo {pages} {405} (\bibinfo {year}
  {2018})}\BibitemShut {NoStop}%
\bibitem [{\citenamefont {Burns}\ \emph {et~al.}(2011)\citenamefont {Burns},
  \citenamefont {Mayagoitia}, \citenamefont {Sumpter},\ and\ \citenamefont
  {Sherrill}}]{SMDensity}%
  \BibitemOpen
  \bibfield  {author} {\bibinfo {author} {\bibfnamefont {L.~A.}\ \bibnamefont
  {Burns}}, \bibinfo {author} {\bibfnamefont {{\'{A}}.~V.}\ \bibnamefont
  {Mayagoitia}}, \bibinfo {author} {\bibfnamefont {B.~G.}\ \bibnamefont
  {Sumpter}},\ and\ \bibinfo {author} {\bibfnamefont {C.~D.}\ \bibnamefont
  {Sherrill}},\ }\bibfield  {title} {\bibinfo {title} {Density-functional
  approaches to noncovalent interactions: A comparison of dispersion
  corrections ({DFT-D}), exchange-hole dipole moment ({XDM}) theory, and
  specialized functionals},\ }\href {https://doi.org/10.1063/1.3545971}
  {\bibfield  {journal} {\bibinfo  {journal} {J. Chem. Phys.}\ }\textbf
  {\bibinfo {volume} {134}},\ \bibinfo {pages} {084107} (\bibinfo {year}
  {2011})}\BibitemShut {NoStop}%
\end{thebibliography}%
%

\end{document}